\documentclass[manuscript, screen]{acmart}
\usepackage{booktabs} 

\usepackage[toc,page]{appendix}
\usepackage{rotating}
\usepackage{graphics}
\usepackage{lineno,hyperref}
\usepackage{algorithm}
\usepackage{algorithmic}
\usepackage{amsmath}
\modulolinenumbers[5]
\usepackage{array,multirow}
\usepackage{tablefootnote}

\acmJournal{TKDD}
\acmVolume{9}
\acmNumber{4}
\acmArticle{39}
\acmYear{2017}
\acmMonth{11}

\setcopyright{acmcopyright}

\acmDOI{0000001.0000001}

\begin{document}
\title{SemRe-Rank: Improving Automatic Term Extraction By Incorporating Semantic Relatedness With Personalised PageRank} 

\author{Ziqi Zhang}
\authornote{Corresponding author. The work was carried out while this author was at Nottingham Trent University, UK}
\orcid{0000-0002-8587-8618}
\affiliation{%
  \institution{Information School, University of Sheffield}
  \streetaddress{211 Portobello, Regent Court}
  \city{Sheffield}
  \postcode{S1 4DP}
  \country{UK}}
\email{ziqi.zhang@sheffield.ac.uk}
\author{Jie Gao}
\affiliation{%
  \institution{Department of Computer Science, University of Sheffield}
  \streetaddress{211 Portobello, Regent Court}
  \city{Sheffield}
  \postcode{S1 4DP}
  \country{UK}}
\email{j.gao@sheffield.ac.uk}
\author{Fabio Ciravegna}
\affiliation{%
  \institution{Department of Computer Science, University of Sheffield}
  \streetaddress{211 Portobello, Regent Court}
  \city{Sheffield}
  \postcode{S1 4DP}
  \country{UK}}
\email{f.ciravegna@sheffield.ac.uk}

\begin{abstract}
Automatic Term Extraction deals with the extraction of terminology from a domain specific corpus, and has long been an established research area in data and knowledge acquisition. ATE remains a challenging task as it is known that there is no existing ATE methods that can consistently outperform others in any domain. This work adopts a refreshed perspective to this problem: instead of searching for such a `one-size-fit-all' solution that may never exist, we propose to develop generic methods to `enhance' existing ATE methods. We introduce SemRe-Rank, the first method based on this principle, to incorporate semantic relatedness - an often overlooked venue - into an existing ATE method to further improve its performance. SemRe-Rank incorporates word embeddings into a personalised PageRank process to compute `semantic importance' scores for candidate terms from a graph of semantically related words (nodes), which are then used to revise the scores of candidate terms computed by a base ATE algorithm. Extensively evaluated with 13 state-of-the-art base ATE methods on four datasets of diverse nature, it is shown to have achieved widespread improvement over all base methods and across all datasets, with up to 15 percentage points when measured by the Precision in the top ranked $K$ candidate terms (the average for a set of $K$'s), or up to 28 percentage points in F1 measured at a $K$ that equals to the expected real terms in the candidates (F1 in short). Compared to an alternative approach built on the well-known TextRank algorithm, SemRe-Rank can potentially outperform by up to 8 points in Precision at top $K$, or up to 17 points in F1.
\end{abstract}

%
%

%
%


\maketitle

\section{Introduction}\label{sec:introduction}
Automatic Term Extraction (or Recognition) deals with the extraction of \textit{terms} - words and collocations representing domain-specific concepts - from a collection of domain-specific, usually unstructured texts. It is a fundamental task for data and knowledge acquisition, often a pre-processing step for many complex Natural Language Processing (NLP) tasks. These can include, for example, information retrieval \cite{Lingpeng2005}, cold Start knowledge base population \cite{ellis2015overview,zhang2015lodie}, ontology engineering and learning \cite{Biemann2014,Brewster2007,wong2007tree}, topic detection \cite{el2014scalable,borner2003}, glossary construction \cite{habert1998extending,peng2004chinese,maldonadoself2016}, text summarisation \cite{Mihalcea2004}, machine translation \cite{bowker2003}, knowledge visualisation  \cite{borner2003, blei2009visualizing, chang2009reading}, and ultimately enabling business intelligence \cite{maynard2007,schoemaker2013integrating,palomino2013evaluating}. 

ATE is still considered an unsolved problem \cite{Astrakhantsev2016}, and new methods have been developed over the years to cope with the increasing demand for automated sense-making of the ever-growing number of specialised documentation in industrial, governmental archives and digital libraries \cite{Astrakhantsev2014a,Astrakhantsev2015a,Bordea2013,Bourigault1992,Ananiadou1994,Church1995,Ahmad1999,Frantzi2000,Li2013,Park2002,Penas2001,Matsuo2003,Sclano2007,Rose2010,lossio2014yet,Spasic2013}. These methods typically start with extracting candidate terms (e.g., nouns, noun phrases, or n-grams) using \textit{linguistic processors}, then apply certain \textit{statistical measures} to score the candidates by features collected both locally (surrounding context or document) and globally (typically corpus-level). The scored candidate terms will be ranked for subsequent selection and filtering. 



Although a plethora of methods have been introduced, we notice two limitations of state-of-the-art. \textbf{First}, it is known that no method can consistently perform well in all situations. Comparative studies \cite{Astrakhantsev2016,Zhang2008} have shown that depending on the domains and datasets, the best performing ATE method always varies and the accuracy obtainable by different methods can differ significantly. As a result, knowing and choosing the best performing ATE method a-priori for every situation is infeasible. For this reason, we argue that, instead of aiming to develop an unrealistic `one-size-fit-all' ATE method for any domain, it can be very useful to develop generic methods that when coupled with an existing ATE method, can potentially improve its performance in any domain. The intuition is that, although it can be infeasible to select a-priori the best performing ATE method for a domain, it can be beneficial to know that by applying this `enhancement' to an existing ATE method, we can potentially do better in that domain with this method.

\textbf{Second}, while state-of-the-art typically make use of features such as word statistics (e.g., frequency) to score candidate terms, they often overlook the role of \textit{semantic relatedness}, an important area of research where a significant amount of work has been undertaken over the years, particularly its application in biomedical domain \cite{Agirre2009,Batet2011,cucerzan2007,Lin1998,Strube2006}. Semantic relatedness describes the strength of the semantic association between two concepts or their lexical forms by encompassing a variety of relations between them. A more specific kind of semantic relatedness is \textit{semantic similarity}, where the sense of relatedness is quantified by the `degree of synonymy' \cite{Weeds2003}. For example, \textit{cat} is similar to \textit{dog}, and is related but not similar to \textit{fur}. 
To illustrate the usefulness of semantic relatedness in the context of ATE, assuming \textit{protein} a representative term in a biomedical corpus, then the scores of words highly related to it such as \textit{polymer} and \textit{nitrogenous} should be boosted according to their degree of relatedness with \textit{protein}, in addition to their frequency.


In this work, we introduce \textbf{SemRe-Rank}, the first generic method based on the principle of enhancing existing ATE methods by incorporating semantic relatedness in the scoring and ranking of candidate terms. SemRe-Rank applies a personalised PageRank process \cite{Haveliwala2003} to a semantic relatedness graph of words constructed using word embedding models \cite{Mikolov2013} trained on domain-specific corpus. The PageRank algorithm \cite{Page98} is well-known for its use in computing importance of nodes in a graph based on the links among them, and was originally used to rank webpages. The personalised PageRank extends it by implementing a `bias' (personalisation) in the computation to favour nodes that are more strongly connected to a set of seed (or `starting') nodes. SemRe-Rank \textit{differs from previously related work} in: 1) the way the graph is constructed, and 2) the fact that we use `personalised' PageRank to let a small set of seed nodes to propagate domain knowledge through the graph, eventually helping boost the scoring of real terms. 
Specifically, SemRe-Rank computes a score denoting a notion of `semantic importance' for every word (node) on a graph by aggregating its relatedness with other words on the graph. This is then used to revise the score of a candidate term computed by an ATE algorithm, to obtain a final score. To personalise the PageRank process, we only require the selection of between a dozen and around a hundred real terms through a guided annotation process, and therefore we say that SemRe-Rank is weakly supervised. However, SemRe-Rank can also be completely unsupervised as we demonstrate its robustness in our experiments. 

SemRe-Rank is extensively evaluated with 13 state-of-the-art ATE algorithms on four datasets of diverse nature, and has shown to effectively enhance ATE methods that are based on word statistics as it has achieved widespread improvement over all methods and across all datasets. On many cases, this improvement can be quite significant ($\geq 4$ percentage points), including a maximum of 15 points in terms of the average Precision in the top ranked $K$ candidate terms for a set of $K$'s, and 28 points in terms of F1 measured at a $K$ that equals to the expected real terms in the candidates. Compared to an alternative approach that adapts the well-known TextRank algorithm, SemRe-Rank can potentially outperform by up to 8 points in the Precision at top $K$, or up to 17 points in F1.

Our unique contributions are three-fold. \textbf{Conceptually}, we propose a novel perspective towards the task of ATE and take a previously unexplored venue of research. From the \textbf{methodological} point of view, we introduce a generic method to enhance existing ATE methods by incorporating semantic relatedness in a novel way. \textbf{Empirically}, we undertake extensive evaluation to show that our proposed method can improve a wide range of ATE methods, often quite significantly.

The remainder of this paper is structured as follows. Section \ref{sec:related_work} introduces ATE in details and reviews related work. Section \ref{sec:methodology} describes the proposed method. Section \ref{dataset} describes datasets used for evaluating SemRe-Rank, while Section \ref{exp} presents experiments and evaluation of SemRe-Rank. Section \ref{limit} discusses the limitations of SemRe-Rank, followed by Section \ref{conclusion} that concludes this work and discusses future work.



\section{Related work}\label{sec:related_work}
\subsection{Automatic Term Extraction}
A typical ATE method consists of two sub-processes: \textit{extracting candidate terms} using linguistic processors and statistical heuristics, followed by \textit{candidate ranking and selection (i.e., filtering)} using algorithms that exploit word statistics. \textit{Linguistic processors} often make use of domain specific lexico-syntactic patterns to capture term formation and collocation. They often take two forms: `closed filters' \cite{arora2014improving} focus on precision and are usually restricted to nouns or noun sequences. `Open filters' \cite{Frantzi2000,Aker14} are more permissive and often allow adjectives, adverbs, etc. Both may use techniques including Part-of-Speech (PoS) tag sequence matching, n-gram extraction, Noun Phrase (NP) Chunking, and dictionary lookup. Most often, candidate terms are normalised (e.g., lemmatisation) to reduce inflectional forms and stop words are removed. Simple statistical criteria such as minimal frequency of occurrence may be used to remove candidates that are almost impossible to be terms. Qualified candidate terms can be a simple form, such as `cell' from the biomedical domain, or a complex form consisting of multiple words \footnote{Note that a term can also consist of symbols and digits. However, for the sake of simplicity we refer them universally as `words`.}, such as `CD45RA+ cell' and `acoustic edge-detection' from the computer science domain.

\textit{Candidate ranking and selection} then computes scores for candidate terms to indicate their likelihood of being a term in the domain, and classifies the candidates into terms and non-terms based on the scores. The ranking algorithms are considered the most important and complicated process in an ATE method \cite{kageura1996methods,Astrakhantsev2016} as they are often how an ATE method distinguishes itself from others. The selection of terms are often based on heuristics such as a score threshold, or a section of the top ranked candidate terms \cite{Zhang2016}. In the following, we will focus on candidate ranking algorithms adopted by different ATE methods.

The ranking algorithms usually base on two principles \cite{kageura1996methods}: \textit{unithood} indicating the collocation strength of units that comprise a single term and \textit{termhood} indicating the association strength of a term to domain concepts. We will discuss related work in the groups of `classic' methods that do not consider semantic relatedness (Section \ref{classic}), against those that employ semantic relatedness in measuring termhood (Section \ref{semrel}). While most state-of-the-art ATE methods are unsupervised, recent years have seen an increasing number of machine learning based ATE methods, which often cross the boundaries of traditional ATE categories. For these we discuss them in Section \ref{ml}. Since the majority of literature has been well summarised in previous surveys, here we focus on the hypothesis and principles of these methods.

\subsubsection{Classic unithood and termhood based methods}\label{classic}

\paragraph{\textbf{Unithood.}} This measures collocation strength, hence by definition, it is a type of measure for multi-word terms (\textbf{MWTs}). The fundamental hypothesis is that if a sequence of words occurs more frequently together than chance, it is more likely to be an integral unit and therefore a valid term. A vast number of word association measures fall under this category, such as $z$-test \cite{Dennis1965}, $t$-test \cite{Church1991}, $\chi^{2}$ test and log-likelihood \cite{Dunning1993}, and mutual information \cite{Church1990}. Other recent studies focusing on unithood include that of \cite{Deane2005,Matsuo2003,bouma2009normalized,song2011multi,chaudhari2011lexical,el2014scalable,liu2015mining}. For example, Matsuo et al. \cite{Matsuo2003} firstly rank candidate terms by their frequency in the corpus and a subset (typically top n\%) is selected - to be called `frequent terms'. Next, candidates are scored based on the degree to which their co-occurrence with these frequent terms are biased. This is computed using the $\chi^{2}$ test. 

Although unithood plays an indispensable role in ATE, research has shown that the measures on their own are not sufficient to assess validity of a candidate term \cite{Wong2008}, but often needs to combine measures of termhood.

\paragraph{\textbf{Termhood.}} This measures the degree to which a candidate term is specific to the domain, and this is primarily based on statistics such as occurrence frequency. Termhood measures both single-word terms (\textbf{SWTs}) and MWTs. These include, e.g., total (TTF) \cite{Bourigault1992} or average total (ATTF) term frequency in a corpus \cite{Zhang2016}; the adaptation of classic document-specific TFIDF (term frequency, inverse document frequency) used in information retrieval to work at corpus level by replacing term frequency in each document with total frequency in the corpus \cite{Zhang2016}; and Residual-IDF \cite{Church1995} that measures the deviation of the actual IDF score of a word from its `expected' IDF score predicted based on a Poisson distribution. The hypothesis is that such deviation is higher for terms than non-terms.

Several branches of methods have taken different directions to improve the state-of-the-art using frequency-based statistics, including: focusing on MWTs (typically like \textit{CValue}), using contrastive statistics from reference corpora (e.g., \textit{Weirdness}), considering term co-occurrence context (e.g., \textit{NCValue}), and employing topic-modellings. 

\textit{CValue} \cite{Ananiadou1994} observes that real terms in technical domains are often MWTs and usually not used as part of other longer terms (i.e., nested). Frequency based methods are not effective for such terms as 1) nested candidate terms will have at least the same and often higher frequency, and 2) the fact that a longer string appears $n$ times is a lot more important than that of a shorter string. Thus \textit{CValue} computes a score that is based on the frequency of a candidate and its length, then adjusted by the frequency of longer candidates that contain it. If a candidate term is frequently found in longer candidate terms that contain it, it is called a `nested candidate term' and its importance (i.e., CValue score) is reduced. Several more recent methods such as \textit{RAKE} \cite{Rose2010}, \textit{Basic} \cite{Bordea2013}\footnote{This is the baseline method in \cite{Bordea2013}. For the sake of convenience, we follow \cite{Astrakhantsev2016} to call this `\textit{Basic}'.}, and \textit{ComboBasic} \cite{Astrakhantsev2015a} choose to also promote candidate terms that are frequently nested as part of other longer candidates. \textit{RAKE} firstly computes a score for individual words based on two components: one that favours words nested often in longer candidate terms, and one that favours words occurring frequently regardless of the words which they co-occur with. These are computed using properties of nodes on a co-occurrence graph of words. Then it adds up the scores of composing words for a candidate term. \textit{Basic} modifies \textit{CValue} by promoting nested candidate terms, often used for creation of longer terms. While \textit{CValue} and \textit{Basic} were originally designed for extracting MWTs, \textit{ComboBasic} modifies \textit{Basic} method further by allowing customisable parameters that can be tailored either for extracting SWTs or MWTs.

\textit{Weirdness} \cite{Ahmad1999} compares normalised frequency of a candidate term in the target domain-specific corpus with a reference corpus, such as the general-purpose British National Corpus\footnote{\url{http://www.natcorp.ox.ac.uk}}. The idea is that candidates appearing more often in the target corpus are more specific to that corpus and therefore, more likely to be real terms. \textit{Domain pertinence} \cite{Meijer2014} is a simplification of \textit{Weirdness} as it uses un-normalised frequency. \textit{Relevance} \cite{Penas2001} extends \textit{Weirdness} by also taking into account of the number of documents where candidate terms occur. Astrakhantsev \cite{Astrakhantsev2014a} introduces \textit{LinkProbability}, which uses Wikipedia as a reference corpus and normalises the frequency of a candidate term as a hyperlink caption by its total frequency in Wikipedia pages. However, if a candidate does not match any hyperlinks it receives a score of 0.

\textit{NCValue} \cite{Frantzi2000} extends \textit{CValue} by introducing the notion of `term co-occurrence context'. It hypothesises that 1) a domain-specific corpus usually has a list of `important' words that appear in the vicinity of terms; 2) and that candidate terms found in the context of such words should be given higher weight. It thus firstly computes \textit{CValue} of candidate terms in a corpus, then extracts words from the top $n$ to be `contextual words'. Next the \textit{CValue} of any candidate terms found in the context of these contextual words are boosted by its co-occurrence frequency with these words and their weights. 

The method by \cite{Bolshakova2013, Li2013} uses topic-modelling techniques (e.g., clustering, LDA \cite{Blei2003}) to map the domain corpus into a semantic space composed of several topics. Then probability distribution over the topics for words are used to score candidate terms. For example, \cite{Bolshakova2013} adapt \textit{TTF} and \textit{TFIDF} by replacing term frequency in the corpus with its probability in all topics, and document frequency with topic frequency. \cite{Li2013} combine \textit{TTF} with the sum of the probability of composing words over all topics.

\paragraph{Hybrid} Such methods often adopt linear or non-linear combination of unithood and termhood measures. For example, \cite{Wong2008} propose a method where the score of a candidate term is collectively dependent on `domain prevalence' based on the frequency of a candidate in the target domain, `domain tendency' measuring the degree to which a candidate tends to be found more frequently in the target domain than reference domains, and `contextual discriminative weight' comparing a candidate against important contextual words. \textit{GlossEx} \cite{Park2002} linearly combines `domain specificity' (a termhood measure), which normalises the \textit{Weirdness} score by the length (number of words) of a candidate term, with `term cohesion' (a unithood measure) that measures the degree to which the composing words tend to occur together as a candidate other than appearing individually. \textit{TermEx} \cite{Sclano2007}, further extends GlossEx by linearly combining a third component that promotes candidates with an even probability distribution across the documents in the corpus (i.e., those that `gain consensus' among the documents). \cite{lossioventura2014} combine \textit{CValue}, \textit{TFIDF}, with a unithood measure called `insideness' \cite{Loukachevitch2012} that compares search engine page hits returned for exact matches and non-exact matches. Additionally, voting algorithms \cite{Zhang2008} that take (un-)weighted average of scores returned by several measures also belong to this category. 

\subsubsection{Machine learning based methods}\label{ml}
Given training data, machine learning based methods \cite{Astrakhantsev2014a,conrado2013,Turdakov2014,maldonadoself2016} typically transform training instances into a feature space and train a classifier that can be later used for prediction. The features can be linguistic (e.g., PoS pattern, presence of special characters, etc), or statistical or a combination of both, which often utilise scores calculated by statistical ATE metrics \cite{maldonadoself2016,yuan2017supervised}. However, one of the major problems in applying machine learning to ATE is the availability of reliable training data. Semi-supervised and weakly supervised learning based approach have gained increasing attention in recent years to address this issue. For example, positive unlabelled (PU) learning \cite{Astrakhantsev2014a} follows a bootstrapping approach starting with extracting top 100 - 300 candidate terms using \textit{ComboBasic}, then using these candidates as positive examples to induce a classifier using features such as \textit{CValue}, \textit{DomainCoherence}, \textit{Relevance}, etc. \cite{maldonadoself2016} propose an ongoing retraining method that incorporates domain experts' validation into supervised learning loop and iteratively train a classifier with new training data combining manually labelled examples (by validation) and examples labelled by the previously trained model. \cite{judea2014unsupervised} adopt a heuristic-based method to generate positive and negative examples of technical terms in the patent domain for supervised training. \cite{Aker2013} address the task of bi-lingual term extraction, where the goal is to project terms already extracted from a source-language resource to a different, target-language using parallel corpus. In this case, the source-language terms and the parallel corpus are used to train a machine learning model for the target-language.

Although various attempts have been made, the portability of current machine learning based methods due to the cost of creating quality training data is still arguable. Empirically, they do not always outperform unsupervised, even simple ranking methods \cite{Astrakhantsev2016}.

\subsubsection{Semantic relatedness based methods}\label{semrel}
As shown before, the computation of either unithood or termhood heavily relies on word statistics such as frequencies. However, we argue that the use of (co-)occurrence frequency of words as evidence is insufficient. Semantic relatedness could also be a useful type of signal in statistics based ATE methods, and also as features for machine learning based methods. This is overlooked by the majority of state-of-the-art ATE methods. Here we refer to semantic relatedness based ATE as those methods using explicit measures for quantifying semantic relatedness, the range of which is beyond the scope of this work but surveyed in \cite{Zhang2012}. These exclude, for example, approaches that simply employ the frequency of co-occurrence. 

\textit{KeyConceptsRelatedness} (KCR) \cite{Astrakhantsev2014a} selects terms as those semantically related to some knowingly domain-specific concepts. Firstly, top $n$ domain-specific concepts are extracted following an approach similar to \cite{El-Beltagy2010}. This generally selects candidate terms that are at least above a certain frequency threshold, and appear in the first few hundred of words in a document. Then these filtered candidate terms are ranked by their frequency and the top $n$ are selected. Next, for each candidate term, its semantic relatedness with each of the $n$ concepts are computed, and its final score is the average of the top $k$ ($k<n$) similarities. To compute semantic relatedness, the method trains a word embedding model using Wikipedia, and uses the cosine vector similarity metric. The approach adopted here for computing semantic relatedness belong to the research of measuring \textit{distributional similarity} of words \cite{Weeds2003,mikolov2013efficient,bernier2016evaluation} based on large corpus. This is widely used as a computable proxy for lexical semantic relatedness. 

KCR is highly similar to \textit{Domain Coherence} (DC) \cite{Bordea2013} and the method by \cite{Khan2016}. In DC, `key concepts' are replaced with an automatically constructed domain model consisting of words and phrases considered to be `important'. This is built using the \textit{Basic} measure. Then semantic relatedness with highly ranked words from this model is computed using `normalised PMI' (NPMI). In \cite{Khan2016}, a subset of top ranked candidate terms are extracted using \textit{CValue} and \textit{TFIDF}, and semantic relatedness is also computed using cosine vector similarity based on a word embedding model. 

\cite{lossio2014yet} build a graph of candidate terms based on their pair-wise semantic relatedness and argue that the weight of a candidate term depends on the number of neighbours that it has, and the number of neighbours of its neighbours on the graph. This is similar to the principle of \textit{RAKE} \cite{Rose2010}. Mathematically, semantic relatedness is calculated using a dice-coefficient function based on co-occurrence frequency and the term weight is modelled as a log function. 

Methods of \cite{maynard1999term,Maynard1999,Maynard2000,Maynard2008} revise the NCValue method \cite{Frantzi2000} by modifying the calculation of the weights of contextual words (see Section \ref{classic} under `Termhood'). While in NCValue, the weight of a contextual word depends on its co-occurrence frequency with a subset of candidate terms highly ranked by CValue; in this revised method, this weight is computed based on its semantic relatedness with entries in the selected subset of candidate terms. Using the biomedical domain for experiments, semantic relatedness was computed based on the distance between the semantic categories of a contextual word and a candidate term in the hierarchy provided by the UMLS Semantic Network\footnote{https://semanticnetwork.nlm.nih.gov/}, using a method similar to \cite{Sumita1991}.





\subsubsection{Limitations of state of the art}
\textbf{First}, state of the art methods are typically introduced as standalone, competing alternatives, the performance of which are always domain dependent. For example, \cite{Astrakhantsev2016} show that, among 13 state-of-the-art ATE methods, the best performing methods on a computational linguistic dataset only come the last when tested on a biomedical dataset. This is also confirmed in our experiments in Section \ref{exp}. It is unclear whether and how different methods can be combined to enhance each other, and studies in this direction have been limited to the use of `voting' strategies, where given the same list of candidate terms to rank, the scores computed by a range of methods are given different or equal weights, aggregated, and then used to re-rank the candidate terms. However, on the one hand, determining the weights can require prior knowledge of the expected performance of each method on a dataset \cite{Zhang2008}; on the other hand, voting can inherit limitations of different methods, as previous work \cite{Astrakhantsev2016} has shown that on many datasets, the performance of a voting method can be significantly lower ($\geq$10 percentage points) than the best performing, individual methods combined by voting. In contrast, SemRe-Rank is designed as a generic method to enhance existing ATE methods, and our experiments show that it is effective for a wide range of ATE methods in different domains.

\textbf{Second}, SemRe-Rank makes use of semantic relatedness to `boost' the scores of candidate terms relevant to a domain. This is often an overlooked venue in classic unithood and termhood based methods. And compared to semantic relatedness based methods, SemRe-Rank consumes semantic relatedness in a different way, firstly by using the strength of relatedness to create a graph of connected words to which a PageRank process is applied; and secondly by `personalising' the PageRank process using seeds that are expected to `guide' the selection of candidate terms that are truly relevant to the domain. Empirically, we show that it is more effective than, e.g., an alternative approach adapted from the well-known TextRank algorithm \cite{Mihalcea2004} that constructs and represents a relatedness graph in a different way.

\subsection{Keyword(phrase) and topical phrase extraction}\label{keyword}

A different, but closely related area of research to ATE concerns the extraction of keywords or keyphrases - to be referred to as \textbf{keyphrase extraction} - from documents \cite{Witten1999,Turney2000}. Compared to ATE, keyphrase extraction serves different goals and therefore, often uses different techniques. ATE examines terms that need to be representative for the domain and hence corpus-level (global) features are important to provide comprehensive representation of candidate terms. This is particularly important for, e.g., developing lexical or ontological resources for a domain. Keyphrase extraction on the other hand, treats each document differently and most methods do not consider global information across the whole corpus. Their goal is often to identify a handful of representative keyphrases for document indexing \cite{Turney2000}. 

For this reason, keyphrase extraction often utilises statistics gathered specifically for individual documents, such as the classic TFIDF measure \cite{Witten1999}. A well-known method is TextRank \cite{Mihalcea2004}, which also uses the PageRank algorithm. TextRank builds an undirected and unweighted graph to represent word co-occurrence relations from each document based on a context window, then applies PageRank to compute scores for each word node on the graph. The scores are then used to extract keyphrases for each document.

Supervised machine learning methods are also very common in keyphrase extraction. For example, the recent SemEval 2017 initiative\footnote{\url{https://scienceie.github.io/evaluation.html}} has brought renewed attention to this topic. Here it is re-defined as a supervised tagging task, highly relevant to Named Entity Recognition (NER) \cite{Zhang2013a,Zhang2013b,Nadeau2007}. One of the goals is identifying every mention instance of keyphrases in documents. And all the 17 participating systems have overwhelmingly adapted classic NER techniques, often using machine learning models built with training data. 

Another related area of research concerns \textbf{topical phrase extraction from topic models}, where the goal is to mine representative sequences of words (i.e., phrases) to describe topics computed by topic modelling algorithms on a corpus. Again this serves a different goal, but is similar to ATE as it can be considered as a two-step ATE process where the first step mines the topics described in a corpus, and the second identifies representative keyphrases for these topics. In theory, this does however, add additional layers of computation. Since topic modelling is beyond the scope of this work, our discussion in the following focuses on works that use techniques similar to ATE and compares the `phrase extraction' part of these methods with ATE. 

Earlier methods such as \cite{Wang2007,Wallach2006} propose to extract bi-grams from topic models. ATE however, deals with word sequences of variable length, which is unknown a-priori. \cite{Danilevsky2006} firstly extract order-free, variable length of word sets that are frequent patterns found to belong to the same topics, then compute several metrics to rank these frequent patterns. These metrics are designed to favour patterns that are frequent over the entire corpus (frequency), have high frequency concentrated on a single topic (informativeness), have low frequency as being part of longer patterns (completeness), and whose composing words co-occur significantly more often than the expected chance (collocation). Essentially, the first two metrics can be considered as measures of termhood, while the last two can be measures of unithood. \cite{Blei2009} evaluate the likelihood of a word sequence being a valid topical phrase using a permutation test that captures the same principle of unithood. \cite{el2014scalable} follow a similar idea as \cite{Danilevsky2006} while addressing model scalability and complexity. In ranking candidate phrases, their method also relies on frequency and collocation strength, which is measured using a generalisation of the \textit{t}-statistic. The later work by \cite{liu2015mining} extends both \cite{Danilevsky2006} and \cite{el2014scalable} by adding a supervised classification element to use a small labelled dataset to select quality topical phrases.  \cite{ren2017cotype} and \cite{shang2017automated} recently explore the distantly supervised learning technique to leverage largely available but potentially noisy labelled data from existing knowledge bases to further improve the method proposed in \cite{liu2015mining}.



\section{Methodology}\label{sec:methodology}

\begin{figure}[ht!]
  \centering
    \includegraphics[width=120mm]{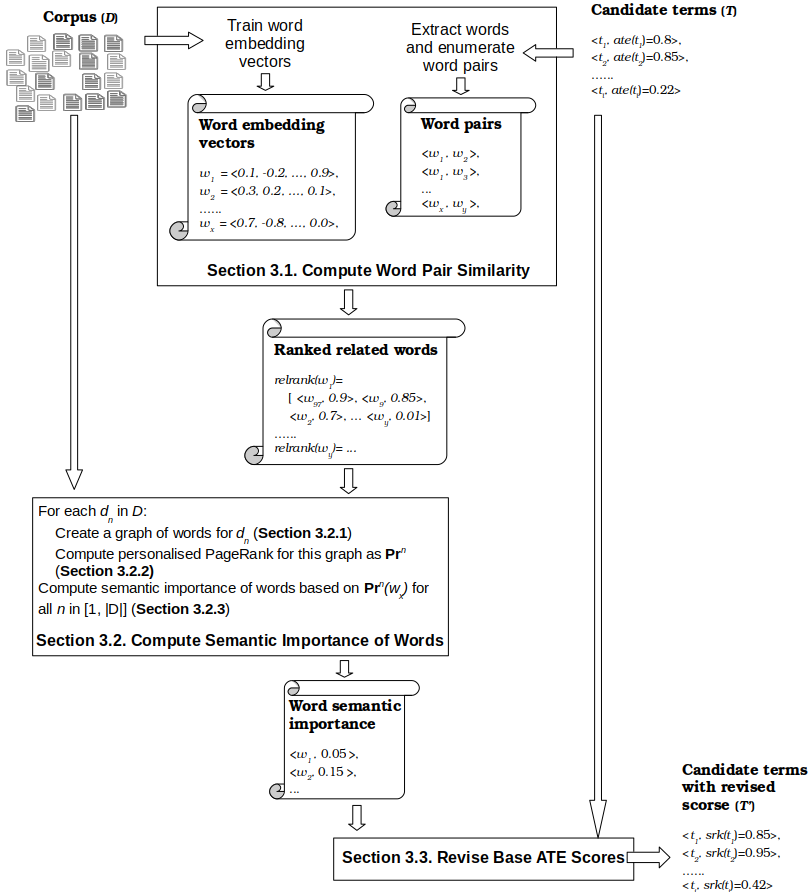}    
      \caption{The overall workflow of the SemRe-Rank method}\label{fig_workflow}
\end{figure}

The workflow of SemRe-Rank is illustrated in Figure \ref{fig_workflow}. The input to SemRe-Rank consists of 1) a target corpus $D = \{d_1, d_2, ..., d_n\}$ from which terms are to be extracted, and 2) a set of candidate terms\footnote{The generation of candidate terms is not the focus of this work, as we use standard approaches depending on different corpus and domains (to be detailed in Section \ref{exp}).} $T = \{t_1, t_2, ..., t_i\}$ that are extracted from $D$ and scored by an existing ATE algorithm (to be called a \textbf{base ATE algorithm}). Also let $ate(t_i)$ denote the score of $t_i$ computed by the base ATE algorithm. The goal of SemRe-Rank is to compute for each candidate term $t_i \in T$, a revised score $srk(t_i)$ by modifying its original ATE score $ate(t_i)$ to incorporate the `semantic importance' of its composing words quantified based on the target corpus.

Let $words(X)$ be a function returning the set of words from $X$\footnote{Also removing stopwords and applying lemmatisation.}, which can be a document $d_n$, a term $t_i$, or a set of candidate terms such as $T$. Starting with $D$ and $T$, we firstly derive the set of words $w_x \in words(T)$ and compute pair-wise semantic relatedness of these words based on the word embeddings trained on $D$ (Section \ref{method_semrel}). Note that we do not use all words from the entire corpus but focus on only words from candidate terms, as we expect them to be more relevant to ATE. Next (Section \ref{method_semantic_importance}), for each document $d_n$, we create a graph for a set of words satisfying $words(d_n) \wedge words(T)$, i.e., the intersection of the words in the document and words from candidate terms extracted for the entire corpus. Words form the nodes on such a graph and edges are created based on their pair-wise semantic relatedness. A personalised PageRank process is then applied to the graph to score the nodes. After applying the process to all documents, for each word $w_x \in words(T)$, we sum up its PageRank score computed within each of its containing document, to derive a `semantic importance' score of the word. This can be considered a quantification of the word's representativeness for the target corpus by incorporating its semantic relatedness with other words in the same corpus. Finally (Section \ref{method_rerank}), for each candidate term $t_i \in T$, we compute a revised score $srk(t_i)$ to take into account both $ate(t_i)$, and the semantic importance of its composing words. This score $srk(t_i)$ then replaces $ate(t_i)$ to be used as the new score to rank candidate terms.

\subsection{Pair-wise semantic relatedness}\label{method_semrel}
We follow the recent methods of using word embedding vectors trained on unlabelled corpus, to compute distributional similarity of words as a proxy for measuring the semantic relatedness of two words \cite{Mikolov2013}. Given the target corpus $D$, we train a word embedding model that maps every unique word in the corpus to a dense vector space of a given dimension, where each dimension represents a latent concept hence each word represented as a probability distribution over a set of latent concepts. Then the semantic relatedness of two words $rel(w_x, w_y)$ is calculated using the cosine function between their vector representations:

\begin{equation} \label{e_rel}
 rel(w_x, w_y) = \frac{ \vec{w_x} \cdot \vec{w_y}}{\lVert \vec{w_x} \lVert \lVert \vec{w_y} \lVert}
 \end{equation}

In the above equation, $\vec{w}$ denotes the vector of the word $w$. While a wide range of methods can be used for computing semantic relatedness of two words \cite{Zhang2012}, comparing their effect on SemRe-Rank is beyond the scope of this work. The benefits of using distributional similarity as proxy for semantic relatedness can be two-fold. First, it potentially avoids out of vocabulary issues. Second, the learned vector representations of words are corpus specific, and potentially can be a better representation of the lexical semantics of words in the target domain than those derived from a general purpose dataset or lexical resources. 

In this work, we use the word2vec \cite{Mikolov2013} algorithm to train word embeddings from unlabelled corpora. word2vec employs a neural network algorithm to learn a dense vector of any arbitrary size for each word in a corpus. Given a target corpus, we apply a pre-process to: 1) remove stop words; 2) lemmatise each word; 3) remove any words that do not contain alpha-numeric characters; and 4) remove any words that contain less than certain number of characters ($minc$) (to be detailed in Section \ref{exp_semrerank_setup} depending on the corpus). The word order is retained. We use the skip-gram variant of the method, known to perform better with small corpus and infrequent words, which is typical for ATE tasks. We use an expected vector dimension of 100, and a context window of 3 for all corpora. The parameter settings are rather arbitrary, as the purpose is solely to create a reasonable model for computing semantic relatedness.

Once we have computed pair-wise relatedness for words in $words(T)$, for each word $w_x \in words(T)$, we rank the list of other words based on their semantic relatedness to $w_x$. These ranked lists will be used for establishing edges on the graph (Section \ref{method_semantic_importance}). Formally, we define $relrank(w_x)$ a function that returns the ranked list of other words for $w_x$: 


\begin{align} \label{e_relrank}
 relrank(w_x) &= (w_1, w_2, w_3, ..., w_l) : y= 1, ...,l \wedge w_y \in \{words(T) - \{w_x\}\} \nonumber \\
&\wedge rel(w_x, w_y) > rel(w_x, w_{y+1})
 \end{align}

\subsection{Computing semantic importance of words}\label{method_semantic_importance}
Here our goal is to use the set of $relrank(w_x)$ computed before to create graph(s) on which we use the personalised PageRank algorithm to compute semantic importance scores of each word. Two design options are available. First, we can create a single graph for the entire corpus and apply the PageRank process to this graph. Second, we can create one graph for each document, applying the PageRank process to each graph, and then aggregate the PageRank scores computed for each word from multiple documents to derive a single score for that word. 

We choose the second approach for two reasons. First, this allows us to capture both local evidence (document-level) as the PageRank process only considers certain words from specific documents; and also global evidence (corpus-level) as the semantic relatedness scores used to establish edges are determined by the embedding representation learned from the entire corpus. Second, from a practical point of view, a document-level graph is much smaller than a corpus-level graph and therefore much more efficient to compute. 


\subsubsection{Graph construction.}\label{sec_graph} Algorithm \ref{alg_graph} illustrates the graph construction process for a document $d_n$. Given the set of candidate terms $T$ and a document $d_n$, we firstly find the intersection of their word sets $words(d_n) \wedge words(T)$. Then for each word $w_x$ in this set, we add a node to the graph (line \ref{two}) and select the \textit{strongly related words} $A_{w_x}$ that is a subset of the intersection (line \ref{three}, $select$). Finally, words in $A_{w_x}$ are added to the graph and an undirected, unweighted edge is created between $w_x$ and every word in $A_{w_x}$ (line \ref{four} onwards).

\begin{algorithm}
\caption{Graph construction}\label{alg_graph}
\begin{algorithmic}[1] 
\STATE Input: $d_n$, $V_n \leftarrow \emptyset$, $E_n \leftarrow \emptyset$
\STATE Output: $G_n=(V_n, E_n)$
\FORALL{$w_x \in \{words(d_n) \wedge words(T)$\}} \label{one}
\STATE $V_n = V_n \cup \{w_x\}$ \label{two}
\STATE $A_{w_x} \leftarrow select(relrank(w_x))$ \label{three}
\FORALL{$w_y \in A_{w_x}$} \label{four}
\STATE $V_n = V_n \cup \{w_y\}$
\STATE $E_n = E_n \cup \{(w_x, w_y)\}$
\ENDFOR
\ENDFOR
\end{algorithmic}
\end{algorithm}

\textit{Strongly related words} are selected based on two thresholds. Given a word $w_x$, their semantic relatedness with $w_x$ must at least pass the minimum threshold $rel_{min}$, and also within the top $rel_{top}$ from $relrank(w_x)$. We set $rel_{min}=0.5$ for the scale of [0, 1.0] and $rel_{top}=15\%$. The values are empirically derived based on a preliminary data analysis detailed in Appendix \ref{appendix_semrerank_thresholds}.

In short, lower $rel_{min}$ can ensure higher connectivity of the graph. We set this to be no less than 0.5, as it is the intuitive middle point of the scale. However, our preliminary analysis shows that the choice of $rel_{min}$ sometimes does not effectively filter unrelated or weakly related words, as we observed that many words can have a semantic relatedness score higher than $rel_{min}$ with almost all other words, regardless of how high $rel_{min}$ is set. This is possibly due to inadequate representations learned from domain-specific corpora \cite{wang2015incorporating,lai2016generate,qasemizadeh2016study}. As a result, this can create many nodes that are directly connected with all other nodes on a graph, which can drastically affect the computation of ranking. As mentioned, increasing $rel_{min}$ did not solve the problem but potentially generates more disconnected components in a graph (in the worst case, many isolated nodes). For this reason, we introduce another threshold $rel_{top}$. \cite{Zhang2016b} have shown in a task of finding equivalent relations from linked data that given a set of relation pair candidates, their degree of relatedness follows a long-tailed distribution and the truly equivalent pairs are those receiving exceptionally high relatedness scores. On average these are around 15\% of the candidate set. We believe this to be a reasonable approximation to our problem and hence assume that, given $relrank(w_x)$, only the top 15\% words from the list can be considered to be `strongly related' to $w_x$. 

While our method filters nodes and edges to be created on a graph, an alternative way would be using the edge weighted PageRank algorithm \cite{Xie2015}, in which case words from the entire vocabulary will be added as nodes and there will be a direct, weighted edge between every pair of nodes on the graph. In theory, this is apparently very inefficient as the graph will be very large and overly dense.  

\subsubsection{Personalised PageRank.}\label{method_pagerank}

Traditionally, PageRank algorithms work with directed graphs. Therefore, we firstly convert the above created undirected graph into a directed one by turning each edge into a pair of opposite directed edges. Then given the directed graph $G=(V, E)$, let $deg(v_x)$ be the out node degree of node $v_x$, $M$ be an $|V| \times |V|$ transition matrix where $M_{y,x} = \frac{1}{deg(v_x)}$ if there is a link from $x$ to $y$, and zero otherwise. Then the personalised PageRank algorithm is formalised as a recursive process until convergence:

\begin{equation} \label{e_pagerank}
 \mathbf{Pr} = cM\mathbf{Pr} + (1-c)\mathbf{v}
 \end{equation}

$\mathbf{Pr}$ is a vector of size $|V|$ where each element is the score assigned to a corresponding node. Initially, this is set to a uniform distribution. $\mathbf{v}$ is a $|V| \times 1$ vector whose elements can be set to bias the computation towards certain nodes, and $c$ is the damping factor that by default, has been set to 0.85. The first term of the sum in the equation models the probability of a surfer reaching any node from a source by following the paths on the graph, while the second term represents the probability of `teleporting' to any node, i.e.,  without following any paths on the graph. 

In the standard PageRank, the vector $\mathbf{v}$ asserts a uniform distribution over all elements thus assigning equal probabilities to all nodes in the graph in case of random jumps. Personalised PageRank however, initialises $\mathbf{v}$ with a non-uniform distribution, assigning higher weights to certain elements considered to be more `important'. We refer to such a $\mathbf{v}$ as \textbf{personalisation vector}. This allows those corresponding nodes to spread their importance along the graph on successive iterations of the algorithm. Effectively, the higher weight of a node makes all the nodes in its vicinity also receive a higher weight. 

We wish to utilise this nature of personalised PageRank to bias the computation of rank scores of nodes on the graph based on some forms of domain knowledge. Intuitively, in an ATE task, if we already know a set of real terms, these can be used as domain knowledge to guide the selection of other terms. However, we have two issues. First, for each document, we have a graph of words instead of terms, which can have multiple words. Second, we are creating one graph for every document, which can be in the multitude of hundreds or thousands in a corpus, and therefore it is infeasible to customise a specific set of seed terms for each document. 

We propose to work around these issues by selecting a set of seed terms \textit{for the entire target corpus} $D$, and then map them to nodes found on each document-level graph. Let $S = \{t_1, t_2, ..., t_s\}$ denote a set of seed terms that are known to be real terms extracted from the target corpus. Then we initialise $\mathbf{v}$ as:

\begin{equation} \label{e_ppagerank}
 \mathbf{v}_x =   
 \begin{cases}
  1 &   w_x \in words(S)\\
  0 &   otherwise
  \end{cases}
 \end{equation}

where $\mathbf{v}_x$ denotes the $x$th element in $\mathbf{v}$, thus also corresponds to the node indexed by $x$ on the graph; $words(S)$ returns a set of words extracted from the set of seed terms $S$. Thus on each document-level graph, only nodes that are found to be part of $words(S)$ are assigned a non-zero weight (to be called \textbf{activated}) in the personalisation vector. Note that the number of these activated nodes can vary depending on individual documents. 


We must ensure $S$ can map to words that are found in individual documents for the personalisation to work. Therefore to create $S$, we propose a guided annotation process, where we firstly select top $z$ most frequent candidate terms extracted from a target corpus, and then manually identify those that are considered as real terms to be used as $S$ for that corpus. Empirically, we ensure $z$ to be reasonably small and therefore, we believe that this level of manual input is not laborious since we only need to verify a small list of candidate terms once for each target corpus. We explain our choice of $z$ in experiments. The reason for focusing on the most frequent list of candidates (hence `guiding' the verification process) is that we expect them to map to also frequent words in the target corpus and therefore, increasing the chance of activating nodes on individual document graphs. 

In theory, this annotation process can be automated in many ways, such as trusting an existing ATE method to rank and select a top section of candidate terms. We discuss these options and empirically explore one possibility of such an unsupervised approach in Section \ref{limit}.

\subsubsection{Semantic importance.} Following the personalised PageRank algorithm, $\mathbf{Pr}$ is computed until convergence, by which point we obtain stable rank scores for all nodes on the graph created for a document. Then the corpus level semantic importance of a word is computed as:

\begin{equation} \label{e_semi}
smi(w_x) = \sum_{d_n \in D}{\mathbf{Pr}^{n}(w_x)}
\end{equation}

$\mathbf{Pr}^{n}(w_x)$ is the rank score for $w_x$ computed on the graph for document $d_n$ (0 if the document does not contain this word). 

\subsection{Revising base ATE scores}\label{method_rerank}
The semantic importance score calculated for each word before is then used to modify the scores of candidate terms computed by a base ATE algorithm. Given the set of candidate terms $T$ extracted and scored by a base ATE algorithm, we firstly normalise each candidate's ATE score by the maximum attained score in the set. We then do the same normalisation to the semantic importance scores of all words in $words(T)$. Then let $nate(t_i)$ and $nsmi(w_x)$ each denote the normalised base ATE score of a candidate term and the normalised semantic importance score of a word, the revised SemRe-Rank score of this term $srk(t_i)$ combines the normalised base ATE score of this term and the normalised semantic importance scores of its composing words as below:






\begin{equation} \label{e_termrankm}
srk(t_i)= (1.0+\frac{\sum_{w_x \in words(t_i)}{nsmi(w_x)}} {|words(t_i)|}) \times nate(t_i)
\end{equation}

\section{Dataset}\label{dataset}
To extensively evaluate SemRe-Rank we compiled four frequently used datasets covering different domains. 

\paragraph{\textbf{GENIA}} The most frequently used dataset in evaluating ATE is the GENIA dataset \cite{kim2003,abulaish2007biological}, a semantically annotated corpus for biomedical text mining. GENIA contains 2,000 Medline abstracts, selected using a PubMed query for the terms \textit{human}, \textit{blood cells}, and \textit{transcription factors}. The corpus is annotated with various levels of linguistic and semantic information. Following \cite{Zhang2016} we extract any text annotated as `cons' (concept) as our list of ground truth terms for this dataset, but exclude `incomplete' terms (e.g., coordinated terms, wildcard terms\footnote{E.g., \textit{CD2 and CD 25 receptors} is a coordinated term as it refers to two terms, \textit{CD2 receptors} and \textit{CD25 receptors}, but the first doesn't appear in the text. For details, see \cite{kim2003}.}).

\paragraph{\textbf{ACLv2}} Recent work by \cite{qzadeh2014,qasemizadeh2016acl} compile a dataset using the publications indexed by the Association for Computational Linguistics (ACL). The dataset consists of two versions, ACL ver1 \cite{qzadeh2014} contains over 10,900 documents, and a list of manually annotated domain-specific terms. Term candidates are firstly extracted by applying a list of patterns based on PoS sequence, and then ranked by several ATE algorithms and the top set of over 82,000 candidates are manually annotated as valid or invalid. The second version ACL ver2 \cite{qasemizadeh2016acl} is a corpus of 300 abstracts from ACL ver1 that are fully annotated for the terminology they contain. Two annotators with expert knowledge in the domain are required to read the abstracts, and follow a detailed set of guidelines to mark lexical boundaries for all the terms they find.

We choose to use the ACL ver2 dataset for a number of reasons. First, the complete ACL ver1 dataset became unavailable at the time of writing as it was replaced by the ACL ver2 dataset\footnote{Following this URL takes us to the web page for ACL ver2, access via \url{https://github.com/languagerecipes/the-acl-rd-tec}. Last retrieved: 15th Jun 2017. }. Second, the annotation exercise was arguably biased, as only highly ranked 82,000 term candidates were annotated, and without access to their original lexical context in the documents. Based on the previous research, this only accounts for 15\% of term candidates extracted using the suggested patterns \cite{Zhang2016}, hence it is likely that a very large proportion of real or correct terms was missed. The ACL ver2 corpus however, was fully annotated in a better controlled way. The original dataset\footnote{\url{https://github.com/languagerecipes/acl-rd-tec-2.0}} was annotated by two annotators. In this work, we simply merge the sets of annotations from the two annotators to create a single list of ground trouth terms for the dataset. In case of conflicts, annotations by the first annotator are used.

\paragraph{\textbf{TTCm} and \textbf{TTCw}} While both GENIA and ACLv2 contain abstracts, we further enrich our dataset collection by adding two corpora containing full-length articles compiled under the TTC (Terminology Extraction, Translation Tools and Comparable Corpora) project\footnote{\url{http://www.ttc-project.eu/}, last accessed on 30th Jun 2017}. The English \textbf{TTC-wind} (TTCw) corpus contains 103 articles for the wind energy domain, while the English \textbf{TTC-mobile}(TTCm) contains 37 articles for the mobile technology domain\footnote{Both datasets originally from: \url{http://www.lina.univ-nantes.fr/?Reference-Term-Lists-of-TTC.html}, last accessed on 30th Jun 2017}. Both corpora are created by crawling the Web and then manually filtered. Ground truth lists of terms for both datasets are also provided. 

In addition, the work by Astrakhantsev \cite{Astrakhantsev2016} also uses a number of other datasets for evaluating ATE. These are not selected for several reasons. Most of these datasets are created for keyword extraction, with documents often having only a handful of keywords as ground truth. Some also contain automatically created ground truth by using a domain thesaurus, which is likely to generate false positives (i.e., items incorrectly labelled as domain specific terms) and false negatives (i.e., items not labelled as domain specific terms but should have been). 

Table \ref{tab_datasets} shows the statistics of all four datasets used in the experiment. The datasets cover different technical domains, various length of documents, and different density of ground truth terms\footnote{All processed forms of these datasets are available at: \url{https://github.com/ziqizhang/data}.}.

\begin{table}[t!]
\centering
\small
\caption{Statistics of datasets used for experiment. \#docs - number of documents in the dataset; \#unique terms - number of unique ground truth terms in each dataset; \#words - number of words (using white space as separator), without any filtering such as stop words removal. Note that this includes duplicates.}\label{tab_datasets} 
\begin{tabular}{l | l |l |l |l |l |l}
\hline
		   &	   &	    &\multicolumn{4}{c}{\#words in docs} \\\cline{4-7}
Dataset    &\#docs	&\#unique terms	&total 	&min	&mean	&max \\
\hline
GENIA     	&2,000  &33,396 	&434,782	&49	&217	&532\\
ACLv2	    &300	&3,059  	&32,182	&10		&107	&300\\
TTCw  	&103	&287  		&801,674	&330	&7,783	&67,088\\
TTCm 	&37		&254  		&304,903	&955	&8,240	&54,727\\
\hline
\end{tabular}
\end{table}

\section{Experiment}\label{exp}

\subsection{Objectives, procedures, and performance measures}\label{exp_overall}
\paragraph{\textbf{Objectives.}}
Our experiments are designed for two objectives. \textbf{First}, we aim to test the capacity of SemRe-Rank as a generic method to improve the performance of existing ATE methods. Thus to prove that the method is generalisable and that results are not by chance, we select a range of 13 state-of-the-art base ATE methods covering different categories. We discuss the selection and evaluation of these base ATE methods in Section \ref{exp_baseline}. \textbf{Second}, we aim to test if SemRe-Rank is a better approach to other alternative, general-purpose methods that can be combined with a base ATE method to improve its performance. For this, we replace SemRe-Rank with a method adapting the well-known TextRank algorithm, i.e., \textbf{adapted TextRank (adp-TextRank)}. We introduce the setup of SemRe-Rank and adp-TextRank in Section \ref{exp_semrerank_textrank_setup}, then apply them to the base ATE methods and compare their effects on improving ATE in Section \ref{exp_semrerank_textrank_result}. 

\paragraph{\textbf{Procedures.}} We firstly run each base ATE method on each dataset discussed before to produce a output list of ranked candidate terms. Next, we add SemRe-Rank and adp-TextRank in turn to the base ATE method to produce a different output list of ranked candidate terms. These output lists are then compared against the lists of real terms compiled from the ground truth, using the performance measures detailed below. 

\paragraph{\textbf{Performance measures.}} We use two measures to evaluate the output from ATE. Precision at $K$ calculates the precision (number of true positives according to the ground truth as a fraction of the number of all candidate terms considered) obtained at rank $K$. This is commonly used for evaluating ATE in previous work \cite{da1999using,Park2002}, and the goal is to assess an ATE method's ability to rank true positives highly. We evaluate different $K$ as (50, 100, 500, 1000, 2000)\footnote{Higher $K$'s such as 3000, 4000 etc are also tested, but results are not very informative for two reasons. First, the ability of almost all ATE methods to rank true positives on top quickly diminishes beyond $K=$2000. Second, for the ACLv2, and the two TTC datasets where the expected true positive terms are around 3000 and less than 300 respectively, increasing $K$ beyond these numbers will certainly include significantly more false positives than true positives. For these reasons, we notice little or no change in P@K beyond 2000 and therefore, we do not report them here.}. For the sake of readability, here we only show the \textbf{average P@K calculated over the five segments}, i.e., \textbf{avg P@K}. 
Detailed results can be found in Appendix \ref{appendix_fr}.

The second measure is inspired by the `R-Precision' used in information retrieval, that is the Precision at the $R$th position in the ranking of results for a query that is expected to have $R$ relevant documents. In this work we propose to calculate Precision (P), Recall (R, number of true positives as a fraction of the number of ground truth), and F1 (harmonic mean of P and R) at a $K$ that equals to the size of the intersection of the extracted candidate terms and the ground truth. In other words, this is the number of expected real terms in the candidates, and we refer to this as the number of `\textbf{Recoverable True Positives}', or \textbf{RTP}. Note that the RTPs of an ATE method may only be a subset of the ground truth for a dataset since no linguistic filters are guaranteed to cover all lexical and syntactic patterns of terms. Also, different ATE methods can use different linguistic filters and therefore, for the same dataset, different ATE methods extract different candidate terms and can have different RTP values. Table \ref{tab_rtp} shows the number of candidate terms and recoverable true positives on each dataset, extracted by each ATE method. Using the GENIA dataset as an example, we calculate P, R, F1 at rank \textit{K=13,831} for the Basic method, and \textit{K=15,603} for the CValue method. Intuitively, a perfect ATE method will obtain 100\% precision and also maximum obtainable recall on that dataset at rank $K$=RTP. 
We will refer to this measure as \textbf{Precision, Recall} and \textbf{F1 at \textit{K}=RTP}, or in short, \textbf{P@RTP, R@RTP, and F1@RTP} (also the F1 mentioned in the abstract and introduction of this article).

\begin{table}[t!]
\centering
\small
\caption{Number of candidate terms extracted by each ATE method on each dataset and their maximum Recoverable True Positives (RTP). The voting method is not included as it uses the output (i.e., same set of candidate terms) from other ATE methods. We use publicly available implementations of these methods and due to the difference in such implementations, it has been impossible to ensure they use identical linguistic filters and extract the identical set of candidate terms. See Section \ref{exp_base_ate} for acronyms of base ATE methods.}\label{tab_rtp} 



\begin{tabular}{l | l |l l |l l}
\hline
\multirow{2}{*}{Dataset}	&\multirow{2}{*}{Ground truth}	&\multicolumn{2}{p{4cm}|}{ATE methods: Basic, ComboBasic, LP, NTM, PU\tablefootnote{Implemented in the ATR4S library and share the same linguistic processors, hence have the same set of candidate terms.}}	&\multicolumn{2}{p{6cm}}{ATE methods: TFIDF, CValue, RAKE, Weirdness, Relevance, GlossEx, $\chi^2$\tablefootnote{Same as above but implemented in the JATE 2.0 library.}} \\\cline{3-6}
					&				&Candidate terms	&RTP			&Candidate terms	&RTP\\ \hline
GENIA	&33,396	&56,704	&13,831	&38,850	&15,603 \\
ACLv2	&3,059	&6,361	&2,090	&5,659	&1,976 \\
TTCw	&287	&59,441	&226	&53,088	&250	\\
TTCm	&254	&35,109	&226	&26,011	&238	\\
\hline

\end{tabular}
\end{table}


\subsection{Implementation}
For all the base ATE methods, we use their existing JATE 2.0 \cite{Zhang2016} and the ATR4S \cite{Astrakhantsev2016} implementations in order to facilitate future comparative studies and reproducibility. The two libraries offer the most comprehensive set of state-of-the-art ATE implementations covering a wide range of different categories of methods. They differ in terms of methods implemented, and also the types of linguistic filters supported. For the set of ATE methods within each library, we use the same linguistic filters for them all. However the two libraries do not support identical linguistic filters, and as a result, methods within each library extract the same set of candidate terms; but the candidate term sets across the two libraries are different. The detailed configurations of these methods can be found in Appendix \ref{appendix_config}. Our implementation of SemRe-Rank is shared online\footnote{\url{https://github.com/ziqizhang/semrerank}}. We run all experiments described below on the same computer with 4 CPU cores and a maximum of 12GB memory. 

\subsection{Evaluation of the base ATE methods} \label{exp_baseline}
As discussed before, to prove that our method is generalisable and our results are not by chance, we select a total of 13 state-of-the-art ATE methods covering different categories of ATE methods detailed below. 

\subsubsection{Selection of base ATE methods}\label{exp_base_ate}

Purely \textbf{unithood based methods} are not often used alone today. Thus we select one method to represent this category: the modified $\chi^{2}$ by \cite{Matsuo2003}.

We choose a total of \textbf{10 termhood based ATE methods} as they represent the majority of state-of-the-art. These include: 
\begin{itemize}  
	\item \textit{using occurrence frequencies\textit{}}: TFIDF \cite{Zhang2008}, which is the most used and also best performing \cite{Zhang2016} compared to other similar variants.
	\item \textit{focusing on MWTs}: CValue \cite{Ananiadou1994}, which is recognised as the most effective method for the biomedical domain, as well as Basic \cite{Bordea2013} and ComboBasic \cite{Astrakhantsev2015a}, both are more recent variants based on CValue; and RAKE \cite{Rose2010}, which computes termhood using graph-based properties.
	\item \textit{using reference corpus}: Weirdness \cite{Ahmad1999} and Relevance\footnote{The original implementation in ATR4S uses frequency of candidate terms in a reference corpus. However, in practice, many terms - particularly MWTs - are not found in the reference corpus, but their composing words. Hence we have adapted the method following the same approach used for Weirdness in \cite{Zhang2008}. The implementation is available at \url{ https://github.com/ziqizhang/jate/tree/semrerank}} \cite{Penas2001} both use frequency of terms observed in a reference corpus; and LinkProbability (LP) \cite{Astrakhantsev2014a}, which uses Wikipedia hyperlink frequencies.
	\item using topic-modelling techniques: Novel Topic Model (NTM) by \cite{Li2013}.
\end{itemize}

For \textbf{hybrid ATE methods} that combine unithood and termhood, we use GlossEx \cite{Park2002}, which has been found to be one of the best performing hybrid methods. We also use a uniform weight voting method (Vote) that, given different rankings of a list of candidate terms calculated by several ATE methods, computes new scores for each candidate term by averaging its ranks from different methods. This is essentially the same as the `weighted voting' \cite{Zhang2008}, except that we use uniform weight for different ATE methods. The reasons are, as discussed before, that on the one hand, the weight for each method requires prior knowledge about its expected performance on each dataset; on the other hand, the benefits of `weighted' voting are not strong as empirically, it can still under-perform its composing methods. We create two versions of the voting method, one aggregates the results of the five ATE methods: Basic, ComboBasic, LP, NTM, and PU (Vote\textsubscript{5}); and the other aggregates the results of the seven ATE methods: TFIDF, CValue, RAKE, Weirdness, Relevance, GlossEx, and $\chi^2$ (Vote\textsubscript{7}). The reason is that the ATE methods within each set have the same candidate term lists, which are required for voting to work.  

For \textbf{machine learning based methods}, we use Positive unlabelled (PU) learning \cite{Astrakhantsev2014a}. 

In addition, we have also tested \textbf{semantic relatedness based methods}, including Key Concept Relatedness (KCR) \cite{Astrakhantsev2014a} and Domain Coherence (DC) \cite{Bordea2013}. Intuitively, it makes little sense to incorporate semantic relatedness into another method based on the same hypothesis, as this will inevitably double-weight semantic relatedness, effectively down-weighting other important features such as word statistics. We have empirically observed evidence which shows that when combined with KCR or DC, SemRe-Rank does not consistently improve their base performance. Therefore practically, we do not recommend using SemRe-Rank with other ATE methods that are also based on the principle of semantic relatedness. 

\subsubsection{Base ATE Results} \label{exp_baseline_eval}

Results for these ATE methods are shown in Tables \ref{tab_baseline_result_avgP} and \ref{tab_baseline_result_f1}. Some may argue that the results of different methods from the two libraries are not directly comparable as they use different sets of candidate terms. However, we believe that this is still useful reference since the highest figures are seen on methods from both libraries, suggesting that the different sets of candidate terms do not bias particular ATE methods. 

\begin{table*}[t!]
\centering
\small
\caption{Average Precision at \textit{K} for the five top segments (50, 100, 500, 1,000, 2,000)  (avg P@K) for the 13 base ATE methods on all four datasets. The highest figures on each dataset under each evaluation metric are in \textbf{bold}. For full results, see Table \ref{tab_baseline_result} in Appendix \ref{appendix_fr}.}\label{tab_baseline_result_avgP} 
\begin{tabular}{|l |l|l |l |l |l|l|l|l|l|l|l|l|l|l|}
\hline
Dataset (avg P@K)     	&\rotatebox[origin=c]{270}{Basic}	&\rotatebox[origin=c]{270}{\shortstack{Combo\\Basic}}	
	&\rotatebox[origin=c]{270}{LP}	&\rotatebox[origin=c]{270}{NTM}	&\rotatebox[origin=c]{270}{PU}	&\rotatebox[origin=c]{270}{Vote\textsubscript{5}}	&\rotatebox[origin=c]{270}{CValue}	  &\rotatebox[origin=c]{270}{\shortstack{Gloss-\\Ex}} 	&\rotatebox[origin=c]{270}{RAKE}	&\rotatebox[origin=c]{270}{\shortstack{Rele-\\vance}}	&\rotatebox[origin=c]{270}{TFIDF}	&\rotatebox[origin=c]{270}{Weirdness}	&\rotatebox[origin=c]{270}{$\chi^{2}$}	&\rotatebox[origin=c]{270}{Vote\textsubscript{7}}\\
\hline
ACLv2	    &.60	&.59	
	&.57	&\textbf{.67}	&.61	&\textbf{.67}	&.60	&.40	&.25	&.38	&.54	&.41	&.47	&.51\\
	
GENIA	    &.65	&.65	
	&.59	&.40	&.65	&.60	&\textbf{.80}	&.66	&.57	&.63	&.72	&.76	&.75	&.69\\
	
TTCm	    &.22	&.22	
	&.01	&.11	&\textbf{.23}	&.20	&.21	&.08	&.00	&.03	&.19	&.08	&.07	&.16\\
	
TTCw	    &\textbf{.24}	&\textbf{.24}	
	&.01	&.06	&.22	&.21	&.23	&.02	&.02	&.00	&.14	&.03	&.12	&.11\\\hline

\end{tabular}
\end{table*}

\begin{table*}[t!]
\centering
\small
\caption{F1 at \textit{K}=RTP for the 13 base ATE methods on all four datasets. The highest figures on each dataset under each evaluation metric are in \textbf{bold}. For full results, see Table \ref{tab_baseline_result} in Appendix \ref{appendix_fr}.}\label{tab_baseline_result_f1} 
\begin{tabular}{|l |l|l |l |l |l|l|l|l|l|l|l|l|l|l|}
\hline
Dataset (F1@RTP)     	&\rotatebox[origin=c]{270}{Basic}	&\rotatebox[origin=c]{270}{\shortstack{Combo\\Basic}}	
	&\rotatebox[origin=c]{270}{LP}	&\rotatebox[origin=c]{270}{NTM}	&\rotatebox[origin=c]{270}{PU}	&\rotatebox[origin=c]{270}{Vote\textsubscript{5}}	&\rotatebox[origin=c]{270}{CValue}	  &\rotatebox[origin=c]{270}{\shortstack{Gloss-\\Ex}} 	&\rotatebox[origin=c]{270}{RAKE}	&\rotatebox[origin=c]{270}{\shortstack{Rele-\\vance}}	&\rotatebox[origin=c]{270}{TFIDF}	&\rotatebox[origin=c]{270}{Weirdness}	&\rotatebox[origin=c]{270}{$\chi^{2}$}	&\rotatebox[origin=c]{270}{Vote\textsubscript{7}}\\
\hline
ACLv2  &.42	&.42	
&.42	&.44	&.43	&\textbf{.49}	&\textbf{.49}	&.41	&.33	&.42	&.48	&.42	&.45	&.47\\
	
GENIA    &.37	  &.38	
&.38	&.41	&.40	&.44	&.45	&.48	&.38	&.49	&.56	&\textbf{.57}	&.51	&.55\\
	
TTCm    &.26	&.26	
&.00	&.13	&.34	&.26	&\textbf{.41}	&.06	&.00	&.04	&.27	&.08	&.27	&.24\\
	
TTCw   &.32	&.32	
&.00	&.12	&\textbf{.34}	&.30	&.30	&.02	&.02	&.00	&.18	&.03	&.13	&.19\\\hline

\end{tabular}
\end{table*}

We notice several patterns from the results. \textbf{First}, neither the supervised machine learning based method nor the voting method consistently outperforms others. The voting method depends too much on its composing methods to perform well and tends to find a `middle ground' of all participating methods, except only a few cases. As a result, it can underperform individual methods. \textbf{Second}, while \cite{Astrakhantsev2016} criticises that many existing works do not compare against more recent methods, it is clear that these methods do not demonstrate consistent advantage over conventional, classic methods, such as CValue, and TFIDF. \textbf{Last but not least}, in line with previous findings \cite{Zhang2008,Astrakhantsev2016,Zhang2016}, no single ATE method can outperform others on all datasets under all evaluation measures. When inspecting P@K for different \textit{K}'s in Table \ref{tab_baseline_result} from Appendix \ref{appendix_fr}, the pattern is stronger as an even larger set of different ATE methods has obtained the best result for different \textit{K}'s. This raises the question of whether a `one-size-fit-all' ATE method is possible, and whether it would be more beneficial to develop methods that can potentially improve a wide range of existing ATE methods. 

The significantly lower performance obtained on the TTCm and TTCw datasets are very much due to the very small amount of ground truth terms compared to relatively large amount of extracted candidate terms (See Table \ref{tab_rtp}). For example, for the Basic method on the TTCw dataset, the RTP is just over 200 and the candidate terms extracted are over 59,000. In other words, we expect the method to rank just over 200 \textit{real} terms highly out of over 59,000 candidates. This is a much more challenging task than, e.g., on the GENIA dataset which has over 13,000 RPT's and over 56,000 candidate terms for the same ATE method. Also, effectively this means that for TTCm and TTCw, the maximum attainable P@K for $K>$RTP will be significantly lower. For example, at \textit{K}=2,000 for TTCm, the maximum attainable precision by this method is only 11\% (0.11) ($\frac{226}{2,000}$). 

Despite the scarcity of real terms in some of the datasets, the significantly varying performance of different ATE methods can be due to the limitation in their hypothesis of what makes a real domain specific term, and hence the method built on that hypothesis. For example, Weirdness promotes candidate terms that contain words found to be `unique' to the target dataset. This is measured by comparing a word's frequency in the target dataset against that in a general purpose corpus. On the GENIA dataset where it obtained the second best avg P@K, it is reasonable to expect that a fair proportion of words in this very technical domain can be quite unique and hence have low frequency in a general purpose corpus. However, in the mobile technology and wind energy domains, a substantial amount of common words such as `frequency', `area', `network', `shaft', `blade', and `wind' are often used as part of domain specific terms. Such words may also have high frequency in the general domain. For this reason, results of Weirdness on the TTCm and TTCw datasets are rather poor. Another example is CValue, which obtained the best result on the GENIA dataset, suggesting that its preference to longer candidate terms over nested, shorter ones works well for this domain. In that case, it would be reasonable to expect Basic and ComboBasic, which modify CValue by also promoting such nested candidate terms, to be less effective. 

Unfortunately, so far we only gain this insight after testing all ATE methods. This raises the question of whether it is possible to develop methods that can assess the `fit' between an ATE method for a corpus a-priori. This may be particularly interesting as it can potentially allow us to predict the optimal ATE methods for a target corpus. However, this is beyond the scope of this work, and will be explored in the future. 

So far we have evaluated the performance of base ATE methods. Next, we add SemRe-Rank or Adp-TextRank to each base ATE method to evaluate their effect on enhancing ATE. 


\subsection{Setup of SemRe-Rank and the Adp-TextRank baseline}\label{exp_semrerank_textrank_setup}
In this section, we describe the configuration of SemRe-Rank and also introduce the Adp-TextRank method which we will use as an alternative baseline to SemRe-Rank for comparison. 

\subsubsection{SemRe-Rank setup.} \label{exp_semrerank_setup}
Following the SemRe-Rank method described in Section \ref{sec:methodology}, we firstly need to build word embedding models that are used to compute pair-wise semantic relatedness between words. Next we need to identify the set of seed terms to initialise the personalisation vectors (Section \ref{method_pagerank}).

\textbf{For the word embedding models}, we follow the method described in Section \ref{method_semrel} to apply the word2vec \cite{Mikolov2013} algorithm\footnote{We use the gensim (\url{https://radimrehurek.com/gensim/models/word2vec.html}) implementation.} to each dataset to train a word embedding model to be used for that dataset. The parameter of the minimum character length of a word ($minc$) is set to be the same as that configured for candidate term extraction described in Appendix \ref{appendix_config}. 

\textbf{For seed term selection}, we aim to select a subset of $z$ most frequent candidate terms in a target dataset for verification. This $z$ must not be too small, in which case we may not be able to identify sufficient true positives (i.e., the seed set of terms $S$) that map to words in every document; it also must not be too large, in which case the manual process can become too laborious. We have tested with $z$=200 and 100, from which we identify a seed set of between 20 and 140 real terms depending on datasets. 
Table \ref{tab_seed_stats} shows the size of the verified seed set of terms for each dataset under different $z$, and the corresponding average number of activated nodes on each document-level graph. Overall, we can see that except the ACLv2 dataset, the verified seed terms only map to a very small number of activated nodes (less than 1\% of all nodes in most cases) on a document-level graph.

\begin{table}[t!]
\centering
\small
\caption{Statistics of seed term selection and graph personalisation for the four datasets. \textit{avg\#nodes}: average number of nodes on a document-level graph; \textit{avg\#nodes activated}: average number of activated nodes in the personalisation vector for each document-level graph; \textit{\#seed terms}: the number of verified seed terms for each dataset. Note that since different ATE methods produce different candidate term lists depending on their implementing libraries (JATE 2.0 or ATR4S), this also impacts on the ranked top frequent candidates as well as the number of nodes on a graph. The table only shows the calculated average figures across all these methods.}\label{tab_seed_stats}
\begin{tabular}{l| l | l |l |l |l}
\hline
\multicolumn{2}{l}{} &ACLv2	    &GENIA	&TTCm	&TTCw\\
\hline
\multicolumn{2}{l|}{avg\#nodes}     				&525  	&2,023	&5,793		&8,813\\
		\hline
\multirow{2}{*}{$z$=200}    &avg\#nodes activated	    &101	&25	&63	&19\\
&		\#seed terms verified	  	&128	&126	&49		&24\\
\hline
\multirow{2}{*}{$z$=100}	&avg\#nodes activated	    &62	&16	&31	&11\\
		&	\#seed terms verified	  	&68	&63	&31		&13\\
\hline
\end{tabular}
\end{table}


\subsubsection{Adp-TextRank baseline.} 
To prove that SemRe-Rank is more effective than alternative approaches, we develop a baseline by modifying the well-known TextRank algorithm. We adapt an existing implementation\footnote{\url{https://github.com/summanlp/textrank}} to also use personalisation benefiting from the same set of seeds identified before to calculate a TextRank score for words within individual document $d_n \in D$. Then we add up the TextRank scores of a given word computed on all documents where the word is found. We call this score `\textbf{corpus level TextRank score}' or \textbf{cTextRank} score of a word. It then replaces our `semantic importance' ($smi(w_x)$) of words, and combines with the base ATE scores of a candidate term in the same way described in Section \ref{method_rerank} to compute a final, revised score.






\subsection{Evaluation of SemRe-Rank and Adp-TextRank}\label{exp_semrerank_textrank_result}

We apply SemRe-Rank and Adp-TextRank with each base ATE method on each dataset to obtain revised rankings of candidate terms. We then evaluate these revised rankings using the same measures described before, and compare these figures against those obtained by the corresponding base ATE method. In the following we firstly analyse SemRe-Rank's results on P@K and F1@RTP in Sections \ref{exp_semrerank_p@k} and \ref{exp_semrerank_f1@rtp}, then discuss a comparison against Adp-TextRank in Section \ref{exp_vs_textrank}. 

\subsubsection{SemRe-Rank improvements in P@K}\label{exp_semrerank_p@k}

\begin{figure}[ht!]
  \centering
    \includegraphics[width=153mm]{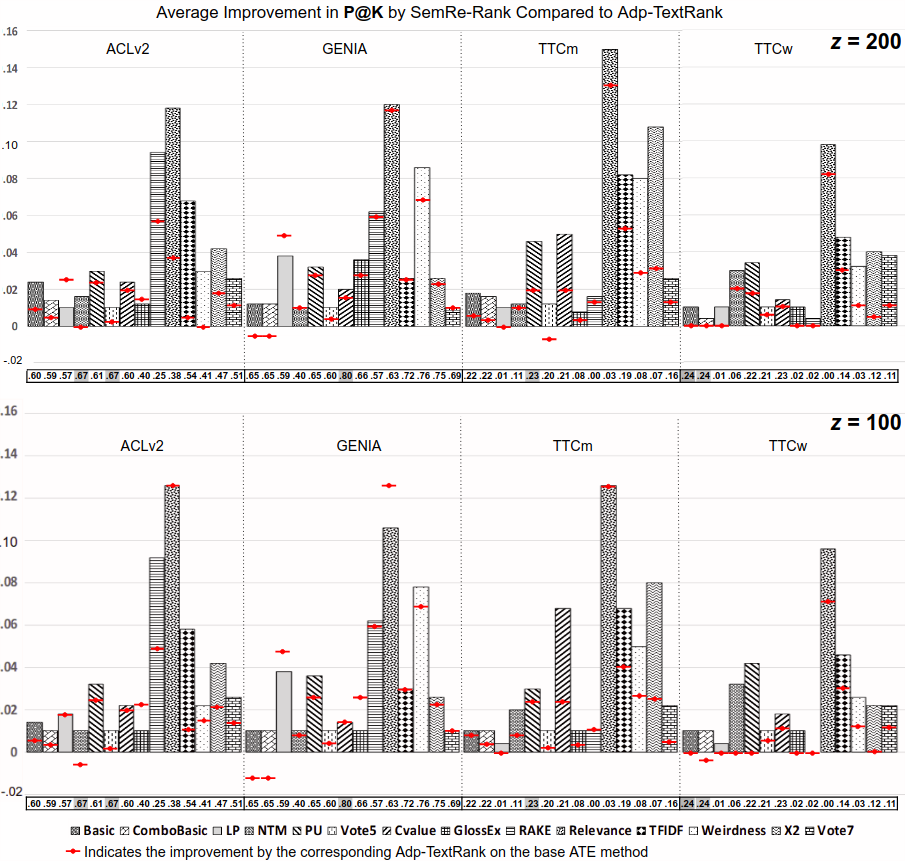}    
      \caption{Comparing SemRe-Rank against Adp-TextRank by \textbf{the improvement in average P@K} over base ATE methods for all five $K$'s considered. The upper graph shows results obtained under $z$=200 and the lower graph under $z$=100. \textbf{Each table column} corresponds to a separate dataset, and contains 14 numbers (with the highest number shaded in grey) corresponding to the average P@K scores obtained by a base ATE method. The order of these base ATE methods shown in the table is the same as that shown in the legend. The base ATE method is also indicated by \textbf{the pattern of the bar} immediately above each number. The \textbf{height of each bar} indicates the improvement by SemRe-Rank over the base ATE's average P@K score shown below it in the table (a missing bar means an improvement of 0). Associated with each column is \textbf{a red line with a dot in the middle}, which indicates the improvement by Adp-TextRank over the same base ATE. For example, the leftmost bar shows that SemRe-Rank improves the Basic algorithm by .024, or 2.4 percentage points (achieving a total of .624, i.e., .60 + .024), in average P@K. Adp-TextRank in comparison, achieves a .01 or 1 precentage point improvement over Basic. (This figure is best viewed in colour) }\label{ssr_avgP}
\end{figure}

We make five observations based on results shown in Figure \ref{ssr_avgP}. \textbf{First}, regardless of the seed size $z$, SemRe-Rank can consistently improve any tested base ATE method in average P@K, with only one exception of RAKE on the TTCw dataset. In the majority of cases, at least 1 percentage point (or .01 on the [0, 1] scale) of improvement is noted. Also in many cases, significant improvements ($\geq$ 4 percentage points) are obtained with different base ATE methods, on all datasets. The maximum improvement is 15 points under $z$=200, or 12.6 under $z$=100. Although there are in total four cases of $<$1 point improvement, considering the wide range of base ATE methods tested, the diverse nature of datasets, also the extreme scarcity of real terms in the TTCm and TTCw datasets, we argue that the task is very challenging and therefore this result is still very promising. It shows that by combining SemRe-Rank with any of the tested and potentially many other ATE methods, in the predominant cases we can expect SemRe-Rank to improve the ATE's capability to rank real terms highly, as measured by P@K. It is worth noting that SemRe-Rank can improve both the best and worst performing base ATE methods on all datasets. On the GENIA dataset, it also significantly improves the second best performing base ATE method Weirdness by 8.6 and 7.8 percentage points under $z$=200 and 100 to obtain an average P@K of .846 and .838 respectively, outperforming the best base ATE CValue+SemRe-Rank (.80+.02 with $z$=200, .80+.014 with $z$=100). The same is noted when comparing CValue against PU on the TTCm dataset under $z$=100. 

\textbf{Second}, relating to Table \ref{tab_seed_stats}, we can see that SemRe-Rank can make effective use of very small amount of domain knowledge in the form of seed terms. With $z$=200, we only identify between 24 and 128 seed terms, and with $z$=100 this drops to only 13 to 68. Notice also that when mapped to activated nodes on document level graphs, on average, only between less than 1\% and 5\% nodes are activated, except on the ACLv2 dataset where this figure is between 10 and 20 \%. As discussed before, in theory, these activated nodes can still contain `noise' because multi-word terms that are selected in the seeds can still contain common words that are not domain-specific.

\textbf{Third}, comparing the results obtained with the two $z$ values, slightly better performance is noticed with $z$=200. However, this is only very noticeable on the TTCm dataset. Again relating to the number of seeds and the activated nodes on a document level graph shown in Table \ref{tab_seed_stats}, it appears that the benefits of having more seed terms - in many cases almost doubled when increasing $z$ from 100 to 200 - are not strong. This can be a desirable feature as it suggests that practically, there is no need for additional human input. 

\textbf{Fourth}, it appears that the base ATE methods that can benefit most from SemRe-Rank regardless of datasets include TFIDF, Weirdness, Relevance, and $\chi^2$. Among these, TFIDF relies on occurrence frequencies and, unlike CValue, Basic etc, does not bias to either SWTs or MWTs. Weirdness and Relevance are based on the hypothetical different frequency distribution of domain specific terms and non-terms. $\chi^2$ relies on candidate term co-occurrences. 

\textbf{Finally}, it is worth noting that since we are calculating the average P@K over five different $K$'s, it is not always the case that we see a change at every \textit{K}. The implication is that, if we exclude the number of \textit{K}'s where no change is noticed, the improvements in P@K can be higher. For details, see Appendix \ref{appendix_fr}.

\subsubsection{SemRe-Rank improvements in F1@RTP}\label{exp_semrerank_f1@rtp}
\begin{figure}[ht!]
  \centering
    \includegraphics[width=153mm]{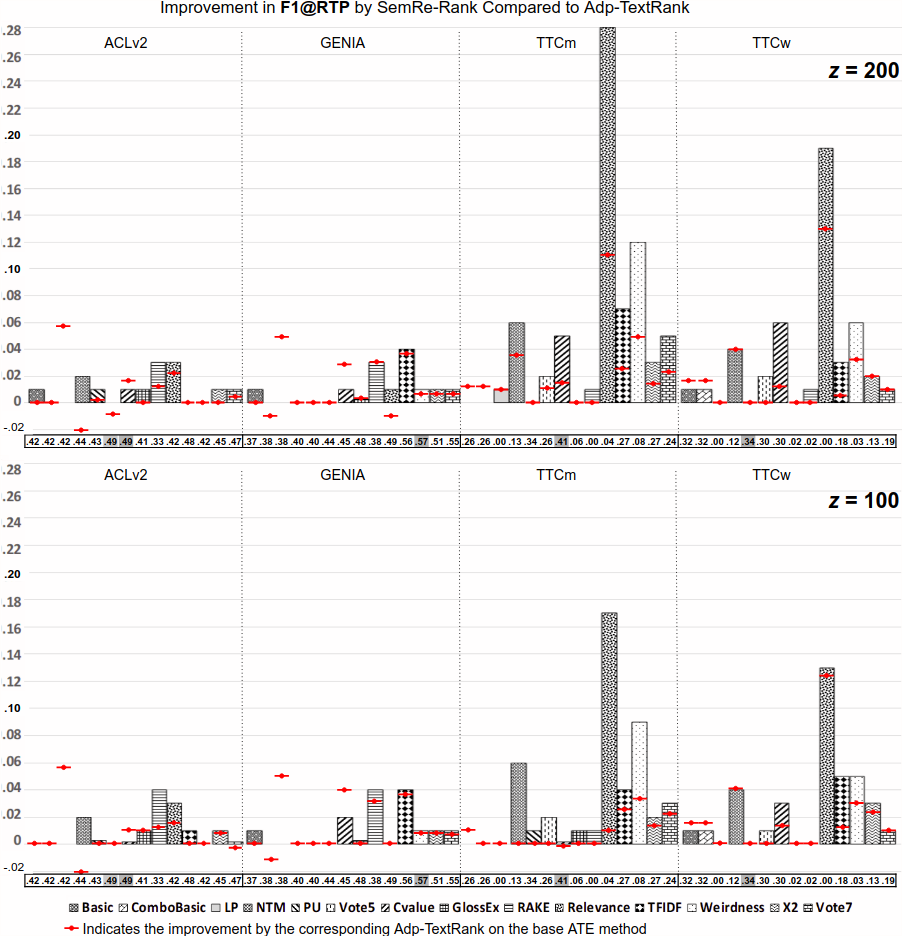}    
      \caption{Comparing SemRe-Rank against Adp-TextRank by \textbf{the improvement in F1@RTP} over base ATE methods. See Figure \ref{ssr_avgP} caption for how to interpret results on this Figure.  (This figure is best viewed in colour) }\label{ssr_f1}
\end{figure}

Figure \ref{ssr_f1} shows that, when measured by F1@RTP, improvements by SemRe-Rank are less noticeable compared to those seen for average P@K, particularly on the ACLv2 and GENIA datasets. This can be attributed to two reasons. First, F1 measures the balance between Precision and Recall. However, on the ACLv2 and GENIA datasets, the maximum attainable Recalls are rather low, due to the low numbers of RTPs compared to the ground truth (see Table \ref{tab_rtp}). Second, on both datasets, P@RTP are likely to be low because the RTP values are higher compared to the $K$'s we have used for evaluating P@K, meaning that we can expect a lot more noise to be in the ranking. The opposite can be said for TTCm and TTCw as in these cases, the RPT values are much lower than the $K$'s we have used to evaluate P@K. Therefore, the achieved improvement in F1@RTP on these datasets are much more significant.

Still we notice many similar patterns as those discussed for P@K. \textbf{First}, using a (potentially very) small number of seed terms, SemRe-Rank effectively improves the ranking of real terms by many base ATE methods, obtaining higher F1@RPT scores. \textbf{Second}, the different improvements achieved under different $z$ values are not very noticeable, except on the TTCm and TTCw datasets. \textbf{Finally}, base ATE methods that have benefited most are also TFIDF, Weirdness, Relevance, and $\chi^2$.  

\subsubsection{SemRe-Rank v.s. Adp-TextRank etc.}\label{exp_vs_textrank}
Compared against Adp-TextRank that uses the same seed sets of terms (both $z$=100 and 200), SemRe-Rank has obtained generally much better performance. Although better results are not \textit{always} achieved for every base ATE method on every dataset, they have been noticed for the most cases, especially in terms of average P@K, and on the TTCm and TTCw datasets where the tasks are more challenging. Specifically, in terms of average P@K, SemRe-Rank can outperform Adp-TextRank by a maximum of around 8 (Relevance, ACLv2) and 6 percentage ($\chi^2$, TTCm) points under $z$=200 and 100 respectively; or in terms of F1@RTP, 17 and 7 points respectively (RAKE, TTCm). Again taking into account the challenges of the tasks due to the wide range of ATE methods and datasets, we argue that the results are rather encouraging. 

One problem with Adp-TextRank is that occasionally, it can damage the performance of base ATE methods, as we notice several cases of drop in both average P@K and F1@RTP. This is a rather unattractive feature, particularly if we cannot anticipate under what situations it will improve or damage base ATE performance. 

Since the key difference between SemRe-Rank and Adp-TextRank is how the graphs are created, we can argue that overall, the superior performance by SemRe-Rank can be attributed to its graph construction approach that may have better captured semantic relatedness between words and subsequently feed that information into the scoring of candidate terms.  

Arguably, the voting method (Vote\textsubscript{5} and Vote\textsubscript{7}) can be seen as another generic approach to improve individual ATE method. Compared to SemRe-Rank, the main problem is that its performance is often limited by the individual best performing method that participates in voting. Tables \ref{tab_baseline_result_avgP} and \ref{tab_baseline_result_f1} have shown that voting cannot always improve the individual best performing method. Previous research \cite{Astrakhantsev2016} has also shown that even weighted voting can still underperform individual participating methods. In contrast, improvements by SemRe-Rank are more consistent, and SemRe-Rank has also proved to be capable of further improving voting based methods (Figures \ref{ssr_avgP} and \ref{ssr_f1}).

\section{Limitations of SemRe-Rank}\label{limit}
In its current state, SemRe-Rank is still limited in a number of ways, which we discuss below and aim to address in our future work.

\subsection{Dependence on supervision}
First and foremost, SemRe-Rank requires a set of seed terms to personalise the PageRank process. Although we have proposed a guided annotation process that helps reduce human input to simply verifying a couple of hundred candidate terms, ideally we want to eliminate this process completely. As discussed before, one method to enable this is to let an existing ATE method to select top ranked $z$ candidate terms and simply use them all to initialise the personalisation vectors. However, due to the varying and unknown performance of ATE methods in different domains, this will inevitably include noise in the personalisation process. To explore if this is feasible, we report our preliminary exploration with some degree of success in this direction. 

To do so, we simply use all top ranked $z$ (either 200 or 100) candidate terms by their total frequency in a corpus. In other words, we remove the human verification process from the current design of SemRe-Rank. Note that although we can test a more sophisticated ATE method and theoretically anticipate better results, our goal here is to gauge the extent to which such a potentially noisy personalisation process will damage the usability of SemRe-Rank as a generic approach to enhance ATE. We will refer to this setting as the unsupervised variant of SemRe-Rank, or simply \textbf{unsupervised SemRe-Rank}. 

\begin{figure}[ht!]
  \centering
    \includegraphics[width=153mm]{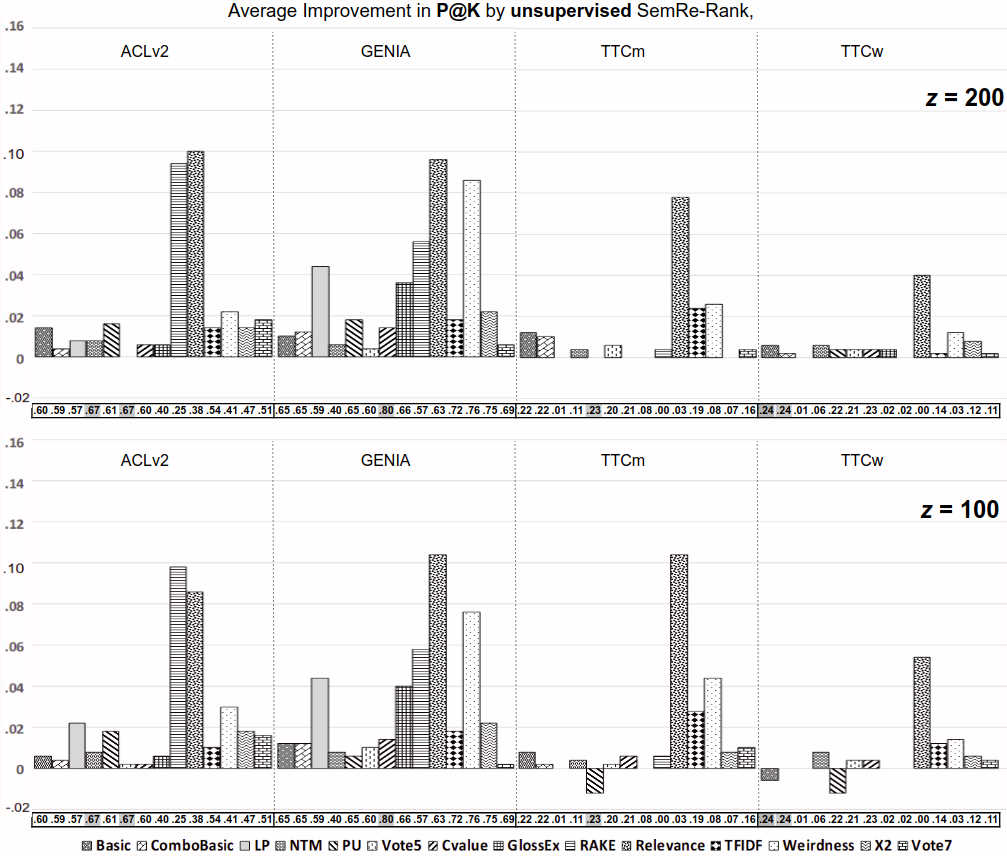}    
      \caption{Improvements in average P@K over base ATE methods by the unsupervised SemRe-Rank. See Figure \ref{ssr_avgP} caption for how to interpret results on this Figure. (This figure is best viewed in colour) }\label{ussr_avgP}
\end{figure}
\begin{figure}[ht!]
  \centering
    \includegraphics[width=153mm]{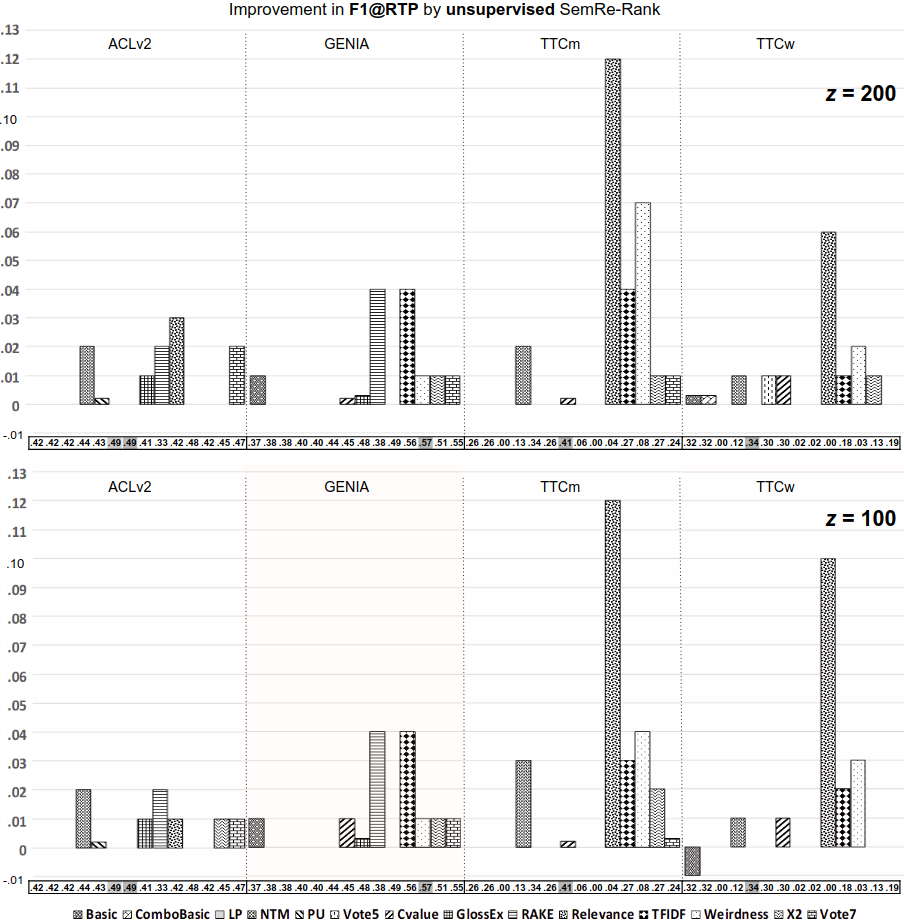}    
      \caption{Improvements in F1@RTP by the unsupervised variant of SemRe-Rank over base ATE methods. See Figure \ref{ssr_avgP} caption for how to interpret results on this Figure.  (This figure is best viewed in colour) }\label{ussr_f1}
\end{figure}

Figures \ref{ussr_avgP} and \ref{ussr_f1} show the improvements in average P@K and F1@RTP over base ATE methods obtained by the unsupervised SemRe-Rank. We summarise three observations from these results. \textbf{First}, compared to the original SemRe-Rank whose results are shown in Figures \ref{ssr_avgP} and \ref{ssr_f1}, the unsupervised variant is indeed less effective, as the ranges of achieved improvements in both measures are lower. This confirms that the noise in the personalisation process indeed has negatively impacted the performance of SemRe-Rank. 

\textbf{Second}, we can see a positive correlation between the amount of noise in seed terms and its negative effect on SemRe-Rank. Recall that Table \ref{tab_seed_stats} shows the number of verified terms for each dataset under different $z$. In other words, the difference between $z$ and the number of verified terms is the number of incorrect, or noisy, candidate terms added to the personalisation process and inevitably, these correspond to poor quality of personalisation vectors, which can mislead the computation of PageRank scores. Specifically, with $z$=200, we have selected 72 incorrect seed terms (or 36\% of all seeds) for ACLv2, 74 (37\%) for GENIA, 151 for TTCm (75\%), and 176 (88\%) for TTCw. The situation is similar with $z$=100, with TTCm and TTCw suffering from a significantly higher proportion of noise. As a result of this, we can see that when compared against the original SemRe-Rank on a per-dataset basis, the performance of unsupervised SemRe-Rank on TTCm and TTCw is significantly lower. 

\textbf{However (our third observation)}, despite the substantial noise in seed terms and their negative effect on the unsupervised SemRe-Rank, it is worth noting that the unsupervised SemRe-Rank has still achieved notable improvements in a wide range of base ATE methods on all datasets. Many of such improvements are also very significant. More interestingly, notice that 1) the noise in seed terms did not cause SemRe-Rank to damage base ATEs, except only three occasions where the decrease is very small; 2) on ACLv2 and GENIA where over 30\% of the seeds are incorrect terms, the performance of the unsupervised SemRe-Rank did not suffer very badly compared to the original SemRe-Rank. This suggests that SemRe-Rank can be quite robust to noise. This is a very important and desirable feature. As in practice, automatically selecting a noise-free seed set of terms is almost impossible. However, creating a seed set with reasonable accuracy but \textit{some} degree of noise is much more achievable. Our results so far have shown SemRe-Rank can potentially still perform just as well using such a reasonable but noisy seed set.

\subsection{Quality of word embeddings}
SemRe-Rank requires learning word embedding vectors on the target corpus in order to compute semantic relatedness between words. Traditionally, word embeddings are best estimated on very large corpora, typically containing multi-million and even billions of words. In comparison, our word embedding learning task is conducted on very small corpora. A known limitation of existing word embedding learning methods is that the embedding vectors of low frequency words are often poor quality \cite{Luong2013}. It is possible that SemRe-Rank can also suffer from this issue, as we did not exclude low frequency words when training word embeddings. To investigate the extent to which rare words can affect SemRe-Rank, we have carried out two further analyses.

\textbf{First}, we aim to understand for a given dataset, the extent to which rare words are used as part (or whole) of real terms. For this we quantify the number of `rare' RTP's found in the candidate terms extracted by each ATE method for each dataset. A rare RTP is one whose composing words are all `rare words'. We call a word `rare' if it has a total corpus frequency below 5, which is the default parameter used in the word2vec implementation to discard any infrequent words. We consider this a minimum requirement for learning `reasonably quality' word embedding vectors. Table \ref{tab_rare_terms} shows that rare RTP's are found in both the ACLv2 and GENIA datasets, but not TTCm or TTCw datasets. Although they represent only a small percentage, this confirms that rare words can potentially impact on SemRe-Rank because they can be used in real terms.

\begin{table}[t!]
\centering
\small
\caption{Number of rare RTPs (Recoverable True Positives) compared to the total number of RTPs found in the candidate term lists of each ATE method. A rare RTP is defined as one whose composing words all have a total corpus frequency of less than 5.}\label{tab_rare_terms} 



\begin{tabular}{l | l l |l l}
\hline
\multirow{2}{*}{Dataset}	&\multicolumn{2}{p{5cm}|}{Basic, ComboBasic, LP, NTM, PU}	&\multicolumn{2}{p{5cm}}{TFIDF, CValue, RAKE, Weirdness, Relevance, GlossEx, $\chi^2$} \\\cline{2-5}
					&Rare RTP	&Total RTP			&Rare RTP	&Total RTP\\ \hline
GENIA	&647	&13,831	&121	&15,603 \\
ACLv2	&143	&2,090	&171	&1,976 \\
TTCw	&0	&226	&0	&250	\\
TTCm	&0	&226	&0	&238	\\
\hline

\end{tabular}
\end{table}

\textbf{Second}, assuming that the embedding vectors of rare words are poor quality, we aim to understand how SemRe-Rank has performed on the RTP's containing these rare words. To do so, we compare the ranking of a rare RTP in the SemRe-Rank's output against that in the base ATE method's output. Specifically, let $rank(ate(t_i))$ return the rank position of $t_i$ among all $T$ candidate terms based on its score computed by a base ATE method, $ate(t_i)$; and let $rank(srk(t_i))$ return the rank position of $t_i$ among the same candidate terms based on its SemRe-Rank revised score ($srk(t_i)$) for this base ATE method. Then we calculate its `relative movement' as:

\begin{equation} \label{e_movement}
mov(t_i) = \frac{rank(ate(t_i))- rank(srk(t_i))}{|T|}
\end{equation}

As an example, if a rare term is ranked at the 999th out of 1,000 candidate terms based on a base ATE method, but the 99th when we apply SemRe-Rank to this base ATE, it will have a movement of $\frac{999-99}{1,000} = 0.90$. In other words, SemRe-Rank has moved this rare term up the entire candidate term list by 90\%. 

\begin{figure}[ht!]
  \centering
    \includegraphics[width=153mm]{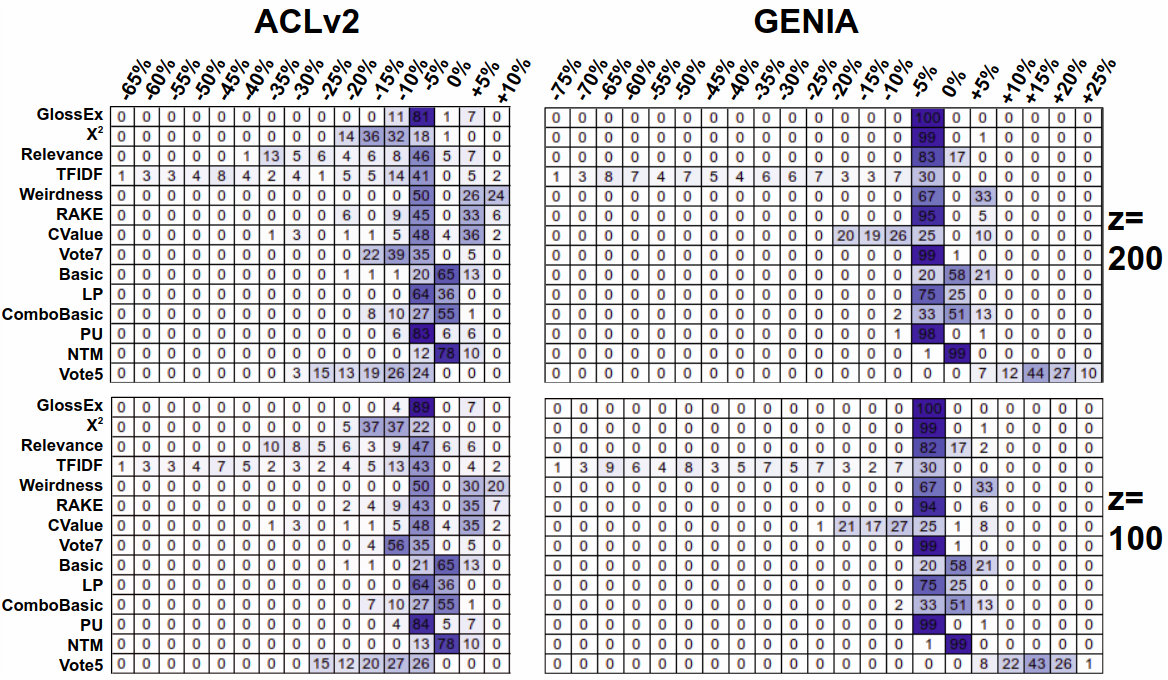}    
      \caption{Heatmap showing the distribution of rare RTPs over different ranges of relative movement in their rankings due to SemRe-Rank, when compared to each base ATE method on either ACLv2 or GENIA dataset. Numbers within each cell are percentage points and each row in a table sums up to 100 (\%). Each column represents a movement range indicated by the percentage numbers on top of the column. Each movement range is a 5\% interval with the maximum indicated by the number, except the 0\% range that represents `no movement' only. For example, in the top left table (ACLv2, $z$=200), the first row indicates that, when we apply SemRe-Rank with $z$=200 to GlossEx, 11\% of rare RTPs are given a new ranking that is down by between 5 and 10 percent compared to their original rankings based on the base GlossEx scores (refer to Table \ref{tab_rare_terms} for the total number of rare RTPs found by each base ATE methods. This figure is best viewed in colour). }\label{heatmap}
\end{figure}

For either of the ACLv2 and the GENIA datasets, and for each base ATE method, we calculate this statistic for every rare RTP found in its candidate terms. We define different ranges of movement based on a 5\% interval on the [-100\%, 100\%] scale (i.e., a movement of between -100\% and -95\%, between -95\% and -90\% etc.), and then we measure the percentage of rare RTP's that fall under each range. Figure \ref{heatmap} plots heatmaps showing the distribution of these rare RTP's over these different movement ranges. It shows that in the majority of cases, SemRe-Rank fails to rank these rare RTP's higher than the base ATE methods. In fact, except those cases of no movement (i.e., 0\%), it has mostly ranked them lower. It is worth noting however, that for those rare RPT's that suffer from up to a 5\% drop in their ranking due to SemRe-Rank, in over 90\% of cases the drop is very minor, i.e., $<1\%$.

These findings show that, although rare RTP's are not common in our datasets, they do cause trouble to SemRe-Rank as it indeed has performed badly on these cases. We further make an assumption that this could be, partly due to the poor embedding vectors estimated for the rare words contained in such rare RTP's. The practical reason for not discarding these rare words when training word embeddings is our need to compute pair-wise relatedness between any words. In this case, we want to have a coverage that is as complete as possible. The relatively small corpus size can certainly be a cause for these poorly estimated embedding vectors. Therefore, as an alternative, we can use already existing word embeddings pre-trained on large general domain corpora, or train word embeddings on additionally collected domain-specific corpora, if these are available.

\subsection{Maximising the benefits of SemRe-Rank}
A natural question by many readers at this point would be when should we use SemRe-Rank and with what ATE methods in order to maximise its benefits. For the first part of this question, our experiments on an extensive set of base ATE methods have shown that SemRe-Rank is highly generic: we can expect it to work with potentially a wide range of different categories of ATE methods that are based on word statistics. However, it should not be used with methods that already use semantic relatedness in any form. 

The second part of this question is a lot harder to answer and would require significant additional work in the future. It also involves answering two sub-questions: 1) how can we predict the optimal base ATE method for a target corpus; and 2) how much improvement can we expect SemRe-Rank to achieve with this method. For 1), as discussed previously in Section \ref{exp_baseline_eval}, we believe that the performance of a base ATE method on a particular dataset can be predicted if we can measure the `fit' between the hypothesis of the ATE method and the characteristics of the target corpus. For example, by measuring the vocabulary overlap between the target corpus and a reference general-purpose corpus, we may be able to gauge the extent to which methods such as Weirdness and Relevance can be effective, as both promote candidate terms that contain words frequently found in the target corpus but not other non-domain corpora. However, developing a generic, systematic method to quantify such a `fit' still requires significant research but can be very beneficial. For 2), previously we have discussed that SemRe-Rank seems to work best with TFIDF, Weirdness, Relevance and $\chi^2$, each in turn representing the categories of ATEs that use simple occurrence frequencies, measure the different frequency distribution of domain specific terms and non-terms, and rely on candidate term co-occurrences. However, it would be too bold to conclude that SemRe-Rank will always work better with any ATE methods from these categories. In fact, we believe that this will depend on many factors, such as whether the base ATE method is a good fit for the target corpus, and whether the method already (either accidentally or purposefully) ranks highly the candidate terms that happen to contain semantically important words (in which case the effect of SemRe-Rank may be small). All these questions will require further investigation to answer.

\subsection{Graph of words v.s. graph of terms}
SemRe-Rank is currently a model based on graphs of words. However, in a typical ATE task, we expect to extract both SWTs and MWTs. This mismatch between the design of SemRe-Rank and the goal of ATE causes several empirical challenges, such as the seed selection and the initialisation of personalisation vectors discussed before. An alternative design would be to develop SemRe-Rank based on graphs of candidate terms, or n-grams (n$>$1). However, this also creates new questions, such as how to learn embeddings for candidate terms and its influence on the shape of created graphs and their subsequent effect on performance.

\section{Conclusion}\label{conclusion}
Automatic Term Extraction is a fundamental task in data and knowledge acquisition and a long established research area for decades. Despite a plethora of methods introduced over the years, it continues to remain challenging and an unsolved task in some domains, as studies (including this one) have shown poor results in some datasets, and inconsistent performance across different domains. 

This work addresses the problem by taking two under-explored research directions: 1) to propose a generic method that can be combined with an existing ATE method to further improve its performance, and 2) to incorporate semantic relatedness in the extraction of domain specific terms. We have developed SemRe-Rank, which applies a personalised PageRank process to semantic relatedness graphs of words to compute their `semantic importance' scores. The scores are then used to revise the base scores of term candidates computed by another ATE algorithm. 

SemRe-Rank has been extensively evaluated with 13 state-of-the-art ATE methods on four datasets of diverse nature, and is shown to be able to improve over all tested methods and across all datasets. Among these, the best performing setting has achieved a maximum improvement of 15 percentage points in P@K, and scored significant improvements ($\geq$ 4 points in P@K) on many base ATE methods on all datasets. 

\textbf{Lessons learned.} \textit{First}, we have shown SemRe-Rank to be a generic approach that can potentially improve various categories of ATE methods, regardless of their base performance, and on a diverse range of datasets. Some of these improvements can be quite significant, even on some very challenging datasets due to their extreme scarcity of real terms. To the best of our knowledge, this is also the first work in such a direction.

\textit{Second}, SemRe-Rank benefits from only a small amount of supervision, in the form of between just 10 and around a hundred seed terms, selected by a manual verification process. 

\textit{Third}, SemRe-Rank is robust to noise, as our preliminary experiments with an unsupervised variant of SemRe-Rank shows that despite the substantial noise in the automatically selected seed terms, the unsupervised variant is still able to obtain widespread improvement over base ATE methods. In many cases, this can be very close to the original SemRe-Rank.

\textit{Last but not least}, our comparison against an alternative method adapted from the well known TextRank algorithm (adp-Textrank) shows that SemRe-Rank can outperform adp-TextRank in many cases and again, sometimes quite significantly. This suggests that our proposed method for incorporating semantic relatedness via a graph model is more effective.

\textbf{Future work.} We will undertake new research to address the limitations of SemRe-Rank discussed before for our future work. \textit{First}, we will explore different methods to automate the seed term selection to develop unsupervised SemRe-Rank. To start, we will test the usage of existing, generally well performing ATE methods for selecting seed terms. Another alternative would be to use existing domain lexicons such as dictionaries and gazetteers that contain words or terms known to be specific to the domain, but not necessarily overlap with the target corpus. We propose to add such words and terms to the graphs and use them as seeds to propagate their influence to other potentially relevant candidate terms found in the corpus. However, this will also require a modification to the word embedding learning process.  

\textit{Second}, we will explore the effects of different word embeddings, including learning embedding vectors from additionally collected large, domain specific corpus, as well as those pre-trained on general purpose corpora. This will help us understand to what extent can we address the issues of rare words and their implications on the performance of SemRe-Rank.

\textit{Third}, we will research methods able to predict optimal ATE methods given a specific target corpus, by measuring a `fit' between the hypothesis of an ATE method and the characteristics of the corpus, such as the way discussed before for Weirdness. We will start with specific ATE methods, then investigate methods for generalisation. Further, additional experiments will be carried out to establish whether SemRe-Rank is particularly effective for certain types of ATE methods.  

\textit{Finally}, we will develop SemRe-Rank on a graph of candidate terms instead of words, and compare its performance against the current implementation based on words.


\bibliographystyle{ACM-Reference-Format}
\bibliography{mybibfile}
\begin{appendices}
\section{Empirical data analysis to determine the $rel_{min}$ and the $rel_{top}$ thresholds} \label{appendix_semrerank_thresholds}
As described in Section \ref{sec_graph}, during graph construction, we need to select `strongly related' words to a target word $w_x$, with which we establish edges on the graph. We use two thresholds to control the selection of such strongly related words for a target word: a minimum semantic relatedness threshold $rel_{min}$, and top $rel_{top}$ from $relrank(w_x)$. This design is empirically driven by a data analysis that is independent from the evaluation of SemRe-Rank. 

We choose to analyse a range of $rel_{min}$ values and their effect on the shape of the created graphs. For this, we have set $rel_{min}$ to be one of the values \{0.5, 0.6, 0.7, 0.8, 0.9\}. \textbf{Firstly}, on each dataset and with each value of $rel_{min}$, we count the number of $w_x \in words(T)$ ($T$ the extracted candidate terms in a dataset) such that for \textit{every other word} $w_y \in words(T) (w_x \neq w_y)$,  $rel(w_x, w_y) < rel_{min}$. In other words, $w_x$ is an isolated node on the graph. We then divide this count by the size of $words(T)$ to obtain a percentage number and show this in Table \ref{tab_disconnected_nodes} for different $rel_{min}$. Note that as discussed before in Section \ref{exp_overall} (last paragraph), the size of $T$ depends on different ATE methods which may use different linguistic filters. And in this work, this depends on either the ATR4S or the JATE 2.0 library that uses its own linguistic filters for the implemented ATE methods. However, we notice the same pattern regardless of what these $T$ are. Therefore, we only discuss our findings in this section based on the $T$ extracted by the ATR4S library.

\begin{table}[t!]
\centering
\small
\caption{Percentage of words that has no strongly related words under a given $rel_{min}$ threshold. These words will become isolated nodes when the graph is constructed for its containing document. }\label{tab_disconnected_nodes} 
\begin{tabular}{l | l |l |l |l |l}
\hline
	&0.9	&0.8	&0.7 	&0.6	&0.5\\
\hline
ACLv2     	&16\%  &9\% 	&6\%	&4\%	&3\%\\
GENIA     	&19\%  &5\% 	&2\%	&0.4\%	&0.1\%\\
TTCm     	&10\%  &4\% 	&3\%	&2\%	&1\%\\
TTCw     	&11\%  &4\% 	&2\%	&1\%	&0.4\%\\
\hline
\end{tabular}
\end{table}

\begin{figure}[ht!]
  \centering
    \includegraphics[width=155mm]{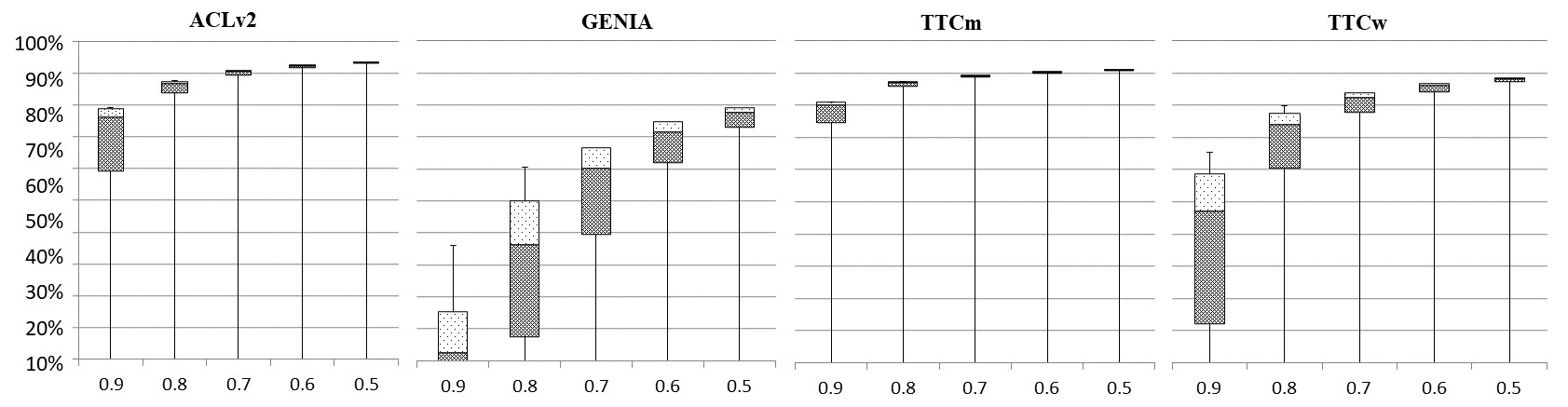}    
      \caption{Distribution of pair-wise semantic relatedness scores computed on the four datasets. \textit{y}-axis: percentage of words from $words(T)$; \textit{x}-axis: $rel_{min}$ threshold.}\label{figure_relmin}
\end{figure}

\textbf{Secondly}, we count for a target word $w_x \in words(T)$, the number of $w_y \in words(T) (w_x \neq w_y)$ such that $rel(w_x, w_y) \geq rel_{min}$. We then divide this number by the size of $words(T)$, obtaining a percentage value showing the fraction of words in $words(T)$ that has a relatedness score of at least $rel_{min}$ with the target word. We call this percentage value `Percentage of Strongly Related Words  (\textbf{PSWA})'. We repeat this for every word in $words(T)$ using the same $rel_{min}$, this gives us a distribution of words from  $words(T)$ over different value ranges of PSWA for a certain $rel_{min}$. We then plot this distribution in quartiles using the box-and-whisker chart in Figure \ref{figure_relmin}, showing for a certain $rel_{min}$ ($x$-axis), the lowest PSWA, the lower quartile, the median, the upper quartile, and the highest PSWA (all referenced against the $y$-axis). 

Using ACLv2 for example, when $rel_{min}=0.9$, the PSWA has a lowest value of 0 and a lower quartile of about 60\%, suggesting that roughly 25\% of words (from $words(T)$, same for the following) have a semantic relatedness score of above 0.9 with between 0 and almost 60\% of other words. The median PSWA is slightly above 75\%, suggesting that about 25\% of words have a relatedness score of above 0.9 with between 60 and 75\% of other words. Or incrementally, 50\% of words (anywhere below the median) can have a semantic relatedness score of above 0.9 with some other words (ranging between 0 and 75\%). Effectively, this means that if we use $rel_{min}=0.9$ as the minimum threshold, almost 50\% of words will be connected with between 60 and almost 80\% of other words on the graph (between the lower and upper quartiles), which seems to make little sense. And yet Table \ref{tab_disconnected_nodes} shows that still for this dataset, 16\% of words are not connected to any other word at all with this threshold, and therefore, become disconnected nodes on a graph. Similar situation is found on the TTCm and TTCw datasets. While on the GENIA dataset, a high $rel_{min}$ does seem to have stronger discriminative power. However, the problems are that, on the one hand, high $rel_{min}$ threshold does not demonstrate consistent discriminating power on all datasets; on the other hand, it almost certainly results in poor graph connectivity as too many nodes are isolated. 

Although reducing $rel_{min}$ certainly creates more superfluous connections, the positive effect is the reduction in the number of isolated nodes from graphs. However, it is clear that $rel_{min}$ alone is insufficient for the task and therefore, we introduce the other threshold $rel_{top}$ to take only the top ranked words from $relrank(w_x)$ for a given $w_x$. And as described before, we set $rel_{min}=0.5$, which although does not eliminate isolated nodes, still reduces them to reasonable levels and semantically represents a middle point on a [0, 1] scale relatedness. And we set $rel_{top}$ to 15\% based on the intuition discussed before in \cite{Zhang2016}.




\section{Full results} \label{appendix_fr}
Table \ref{tab_baseline_result} shows the full results obtained by the 13 base ATE methods. Tables \ref{tab_result_semrerank_full_z=100} and \ref{tab_result_semrerank_full_z=200} show the improvement (or decrease) to the base ATE performance obtained by SemRe-Rank and its unsupervised variant. In both tables, \textbf{avg P@K} is the average of Precision over the five different $K$'s. However, it is not always the case that we notice an improvement in Precision at every $K$. Therefore \textbf{P@K CNGs} shows the number of $K$'s where a change to the base ATE method is noticed. In other words, if we exclude the number of $K$'s where no change is noticed during the calculation of avg P@K, the figures can be higher. 

\begin{table*}[t!]
\centering
\small
\caption{Full result of the 13 base ATE methods on all four datasets. The highest figures on each dataset under each evaluation metric are in \textbf{bold}.}\label{tab_baseline_result} 
\begin{tabular}{|l |l|l |l |l |l|l|l|l|l|l|l|l|l|l|}
\hline
Metric     	&\rotatebox[origin=c]{270}{Basic}	&\rotatebox[origin=c]{270}{\shortstack{Combo\\Basic}}	
	&\rotatebox[origin=c]{270}{LP}	&\rotatebox[origin=c]{270}{NTM}	&\rotatebox[origin=c]{270}{PU}	&\rotatebox[origin=c]{270}{Vote\textsubscript{5}}	&\rotatebox[origin=c]{270}{CValue}	  &\rotatebox[origin=c]{270}{\shortstack{Gloss-\\Ex}} 	&\rotatebox[origin=c]{270}{RAKE}	&\rotatebox[origin=c]{270}{\shortstack{Rele-\\vance}}	&\rotatebox[origin=c]{270}{TFIDF}	&\rotatebox[origin=c]{270}{Weirdness}	&\rotatebox[origin=c]{270}{$\chi^{2}$}	&\rotatebox[origin=c]{270}{Vote\textsubscript{7}}\\
\hline
\multicolumn{15}{|c|}{ACLv2}\\
\hline
P@50	    &.84	&.82	
	&.72	&\textbf{.88}	&.82	&.82	&.62	&.44	&.18	&.32	&.64	&.40	&.58	&.54\\
P@100	    &.72	&.71	
	&.69	&.81	&.82	&\textbf{.85}	&.69	&.46	&.15	&.35	&.65	&.50	&.62	&.46\\
P@500	    &.56	&.55	
	&.56	&\textbf{.67}	&.60	&.63	&\textbf{.67}	&.34	&.29	&.42	&.53	&.36	&.48	&.48\\
P@1,000	    &.49	&.49	
	&.51	&\textbf{.60}	&.43	&.58	&.56	&.36	&.29	&.42	&.47	&.40	&.45	&.46\\
P@2,000	    &.39	&.39	
	&.39	&.41	&.40	&\textbf{.46}	&.45	&.38	&.32	&.40	&.43	&.40	&.41	&.42\\
\hline
P@RTP	&.38	&.38	
	&.39	&.40	&.40	&\textbf{.45}	&\textbf{.45}	&.38	&.32	&.40	&.43	&.39	&.41	&.42\\
R@RTP   &.48	&.47	
	&.47	&.50	&.46	&\textbf{.54}	&\textbf{.54}	&.44	&.35	&.44	&\textbf{.54}	&.44	&.51	&.51\\
F1@RTP  &.42	&.42	
	&.42	&.44	&.43	&.49	&\textbf{.49}	&.41	&.33	&.42	&.48	&.42	&.45	&.47\\
\hline
\multicolumn{15}{|c|}{GENIA}\\
\hline
P@50	    &.80	&.80	
	&.38	&.32	&.74	&.66	&.86	&\textbf{.88}	&.68	&.86	&.68	&.78	&.66	&.82\\
P@100	    &.74	&.74	
	&.51	&.39	&.69	&.58	&\textbf{.83}	&.82	&.63	&.78	&.65	&.74	&.69	&.80\\
P@500	    &.64	&.64	
	&.70	&.42	&.65	&.58	&\textbf{.80}	&.58	&.56	&.58	&.74	&.78	&.71	&.73\\
P@1,000	    &.57	&.57	
	&.69	&.45	&.61	&.60	&\textbf{.78}	&.53	&.52	&.50	&.77	&.77	&.71	&.70\\
P@2,000	    &.49	&.49	
	&.66	&.41	&.58	&.58	&.74	&.47	&.44	&.44	&\textbf{.77}	&.74	&.67	&.70\\
\hline
P@RTP   &.32	  &.33	
	&.34	&.36	&.35	&.39	&.40	&.44	&.36	&.45	&.50	&\textbf{.53}	&.46	&.50\\
R@RTP   &.44  &.44	
	&.43	&.48	&.47	&.51	&.52	&.52	&.41	&.53	&\textbf{.63}	&.62	&.58	&.62\\
F1@RTP    &.37	  &.38	
	&.38	&.41	&.40	&.44	&.45	&.48	&.38	&.49	&.56	&\textbf{.57}	&.51	&.55\\
\hline
\multicolumn{15}{|c|}{TTCm}\\
\hline
P@50	    &\textbf{.52}	&\textbf{.52}	
	&0		&.16	&.44	&.38	&.34	&.20	&0		&0		&.34	&.20	&.18	&.28\\
P@100	    &.35	&.35	
	&0		&.14	&\textbf{.39}	&.33	&.29	&.10	&0		&0		&.29	&.10	&.31	&.20\\
P@500	    &.11	&.11	
	&.01	&.11	&.17	&.13	&\textbf{.21}	&.04	&0		&.03	&.16	&.05	&.13	&.15\\
P@1,000	    &.07	&.07	
	&.01	&.08	&.10	&.09	&\textbf{.12}	&.03	&0		&.04	&.10	&.04	&.10	&.10\\
P@2,000	    &.06	&.06	
	&.01	&.05	&.05	&.06	&\textbf{.07}	&.02	&0		&.03	&\textbf{.07}	&.03	&\textbf{.07}	&\textbf{.07}\\
\hline
P@RTP     &.20	&.20	
	&0	&.10	&.27	&.20	&\textbf{.33}	&.05	&0	&.04	&.22	&.07	&.22	&.20\\
R@RTP     &.36	&.36	
	&0	&.19	&.47	&.36	&\textbf{.55}	&.06	&0	&.05	&.37	&.09	&.36	&.31\\
F1@RTP    &.26	&.26	
	&0	&.13	&.34	&.26	&\textbf{.41}	&.06	&0	&.04	&.27	&.08	&.27	&.24\\
\hline
\multicolumn{15}{|c|}{TTCw}\\
\hline
P@50	    &\textbf{.52}	&\textbf{.52}	
	&0		&0		&.46	&.44	&\textbf{.52}	&.04	&.04	&0		&.26	&0	&.24	&.16\\
P@100	    &\textbf{.41}	&\textbf{.41}	
	&0		&.07	&.34	&.30	&.36	&.04	&.02	&0		&.21	&.04	&.14	&.19\\
P@500	    &.14	&.14	
	&.01	&.10	&\textbf{.16}	&.15	&.15	&.01	&.01	&0		&.10	&.02	&.09	&.09\\
P@1,000	    &.07	&.07	
	&.01	&.07	&\textbf{.09}	&\textbf{.09}	&\textbf{.09}	&.01	&.01	&0		&.07	&.02	&.07	&.07\\
P@2,000	    &.04	&.04	
	&.01	&.04	&\textbf{.05}	&\textbf{.05}	&\textbf{.05}	&.01	&.01	&.01	&\textbf{.05}	&.02	&\textbf{.05}	&\textbf{.05}\\
\hline
P@RTP    &.25	&.25	
	&0	&.09	&\textbf{.26}	&.23	&.23	&.02	&.01	&0	&.14	&.02	&.10	&.14\\
R@RTP    &.44	&.44	
	&0	&.19	&\textbf{.50}	&.43	&.43	&.02	&.03	&0	&.25	&.04	&.20	&.28\\
F1@RTP   &.32	&.32	
	&0	&.12	&\textbf{.34}	&.30	&.30	&.02	&.02	&0	&.18	&.03	&.13	&.19\\
\hline
\end{tabular}
\end{table*}

\begin{table*}[t!]
\centering
\small
\caption{Comparing SemRe-Rank (SRK) and its unsupervised variant (uSRK, both with $z=100$) against each base ATE method. Only the changes over the base ATE methods are shown as points within a scale of [0, 1], and (brackets) indicate negative changes. \textbf{Bold} texts highlight the higher (if different) value between SRK and uSRK on each compared metric. }\label{tab_result_semrerank_full_z=100} 
\resizebox{150mm}{!}{
\begin{tabular}{|l |l |l |l|l |l|l|l|l|l|l|l|l|l|l|}
\hline
Metric     	&\rotatebox[origin=c]{270}{Basic}	&\rotatebox[origin=c]{270}{\shortstack{Combo\\Basic}}		&\rotatebox[origin=c]{270}{LP}				&\rotatebox[origin=c]{270}{NTM}	&\rotatebox[origin=c]{270}{PU}	&\rotatebox[origin=c]{270}{Vote\textsubscript{5}}			&\rotatebox[origin=c]{270}{CValue}	  &\rotatebox[origin=c]{270}{\shortstack{Gloss-\\Ex}} 	&\rotatebox[origin=c]{270}{RAKE}	&\rotatebox[origin=c]{270}{\shortstack{Rele-\\vance}}	&\rotatebox[origin=c]{270}{TFIDF}	&\rotatebox[origin=c]{270}{\shortstack{Weird-\\ness}}	&\rotatebox[origin=c]{270}{$\chi^{2}$}	&\rotatebox[origin=c]{270}{Vote\textsubscript{7}}\\
\hline
\multicolumn{15}{|c|}{ACLv2}\\
\hline
SRK P@K CNGs  	&\textbf{4}	&3		&3		&2			&\textbf{5}	 &\textbf{4}	&4				&\textbf{5}	&5		&5				&\textbf{4}		&4		&\textbf{5}		&4\\
uSRK P@K CNGs  	&1			&3		&3		&\textbf{4}		&4		&3		&4		&4		&5		&5		&3		&4		&3		&4\\

SRK avg P@K		&\textbf{.014}		&.01	&.018			&.01	&\textbf{.032}	&\textbf{.01}	&\textbf{.022}	&.01		&.092			&\textbf{.126}	&\textbf{.058}	&.022			&\textbf{.042}	&\textbf{.026}\\
uSRK avg P@K	&.01				&.01	&\textbf{.022}	&.01	&.018			&.004			&.004			&.01		&\textbf{.098}	&.086			&.01			&\textbf{.03}	&.018			&.016\\
\hline

SRK P@RTP   &-		&-		&-	&-		&-				&-		&\textbf{.01}	&.01	&\textbf{.03}		&.01			&\textbf{.01}		&-	&.01	&.01\\
uSRK P@RTP    	&-		&-	&-	&-		&-		&-	&-	&.01	&.02	&.01	&-		&-	&.01	&.01\\
SRK R@RTP   &.003	&-		&-	&.04	&\textbf{.01}	&-		&\textbf{.002}	&.01	&\textbf{.04}		&\textbf{.03}	&\textbf{.01}		&-	&.01	&.002\\
uSRK R@RTP    	&.003	&-	&-	&.04	&.005	&-	&-	&.01	&.02	&.02	&.002	&-	&.01	&.002\\
SRK F1@RTP  &-		&-		&-	&.02	&\textbf{.003}	&-		&\textbf{.01}	&.01	&\textbf{.03}		&\textbf{.02}	&\textbf{.01}		&-	&.01	&.01\\
uSRK F1@RTP   	&-		&-	&-	&.02	&.002	&-	&-	&.01	&.02	&.01	&-		&-	&.01	&.01\\
\hline
\multicolumn{15}{|c|}{GENIA}\\
\hline
SRK P@K CNGs  	&4		&4		&4			&\textbf{5}		&2				&5		&5				&4		&5				&5				&\textbf{4}		&5		&5		&3\\
uSRK P@K CNGs  	&4		&4		&\textbf{5}		&4		&2		&5		&5		&4		&5		&5		&3		&5		&5		&\textbf{4}\\

SRK avg P@K		&.01			&.01			&.038			&.01	&\textbf{.036}	&.01	&.014	&.04	&\textbf{.062}	&\textbf{.106}	&\textbf{.03}	&\textbf{.078}	&\textbf{.026}	&\textbf{.01}\\
uSRK avg P@K	&\textbf{.012}	&\textbf{.012}	&\textbf{.044}	&.01	&.01			&.01	&.014	&.04	&.058			&.104			&.018			&.076			&.022			&.004\\

\hline
SRK P@RTP    	&.01			&-	&-	&-	&-	&-		&\textbf{.02}	&-		&.03	&-				&.04	&.01	&.01	&.01\\
uSRK P@RTP    	&.01	&-	&-	&-	&-	&-		&.01	&-		&.03	&-	&.04	&.01	&.01	&.01\\
SRK R@RTP    	&\textbf{.01}	&-	&-&-	&-	&.004	&\textbf{.02}	&.007	&.04	&-				&.04	&.01	&.01	&\textbf{.01}\\
uSRK R@RTP    	&.002	&-	&-	&-	&-	&.004	&.01	&.007	&.04	&-	&.04	&.01	&.01	&.004\\
SRK F1@RTP   	&.01	&-	&-	&-	&-	&-				&\textbf{.02}	&.003	&.04	&-				&.04	&.01	&.01	&.01\\
uSRK F1@RTP   	&.01	&-	&-	&-	&-	&-		&.01	&.003	&.04	&-	&.04	&.01	&.01	&.01\\
\hline
\multicolumn{15}{|c|}{TTCm}\\
\hline
SRK P@K CNGs  	&4	&\textbf{4}	&\textbf{1}	&\textbf{5}	&\textbf{2}	&\textbf{4}	&3	&3	&\textbf{3}	&5	&4	&5	&3	&\textbf{4}\\
uSRK P@K CNGs  		&4		&2		&-		&2		&1		&3		&3		&\textbf{4}			&2		&5		&4		&5		&3		&3\\
SRK avg P@K	    &.01	&\textbf{.01}	&\textbf{.004}	&\textbf{.02}	&\textbf{.03}	&\textbf{.01}	&\textbf{.068}	&\textbf{.01}	&.01	&\textbf{.126}	&\textbf{.068}	&\textbf{.05}	&\textbf{.08}	&\textbf{.022}\\
uSRK avg P@K	&.01	&.004			&-				&.01			&(.01)			&.004	&.01	&-							&.01	&.104	&.028	&.044	&.01	&.01\\

\hline
SRK P@RTP    	&-	&-	&-	&\textbf{.05}	&\textbf{.01}	&\textbf{.02}		&-	&\textbf{.01}	&\textbf{.01}	&\textbf{.14}	&\textbf{.03}	&\textbf{.08}	&.01	&\textbf{.02}\\
uSRK P@RTP    	&-		&-		&-		&.02	&-	&-		&-		&-		&-		&.08	&.02	&.04	&.01	&-\\
SRK R@RTP    	&-	&-	&.01	&\textbf{.08}	&\textbf{.01}	&\textbf{.02}		&.01	&\textbf{.01}	&\textbf{.03}	&\textbf{.21}	&\textbf{.06}	&\textbf{.11}	&.03	&\textbf{.05}\\
uSRK R@RTP    	&-		&-		&.01	&.04	&-	&-		&.01	&.004	&.01	&.16	&.04	&.05	&.03	&.01\\
SRK F1@RTP   	&-	&-	&-	&\textbf{.06}	&\textbf{.01}	&\textbf{.02}		&.002	&\textbf{.01}	&\textbf{.01}	&\textbf{.17}	&\textbf{.04}	&\textbf{.09}	&.02	&\textbf{.03}\\
uSRK F1@RTP   	&-		&-		&-		&.03	&-	&-		&.002	&-		&-		&.12	&.03	&.04	&.02	&.003\\

\hline
\multicolumn{15}{|c|}{TTCw}\\
\hline
SRK P@K CNGs  	&2	&2	&\textbf{1}	&\textbf{3}	&\textbf{2}	&\textbf{2}	&\textbf{2}	&\textbf{2}	&-	&\textbf{5}	&\textbf{4}	&\textbf{5}	&\textbf{4}	&3\\
uSRK P@K CNGs	&2			&2		&-	&2		&1			&1		&1		&-	&-	&5		&3		&3		&2		&3\\
SRK avg P@K	    &\textbf{.01}	&\textbf{.01}	&\textbf{.004}	&\textbf{.032}	&\textbf{.042}	&.01	&\textbf{.012}	&\textbf{.01}	&-	&\textbf{.096}	&\textbf{.046}	&\textbf{.026}	&\textbf{.022}	&\textbf{.022}\\
uSRK avg P@K	   	&(.006)	&-	&-	&.01	&(.01)	&.01	&.01	&-	&-	&.054	&.012	&.014	&.01	&.01\\

\hline
SRK P@RTP    	&-	&-		&-	&\textbf{.03}	&-	&\textbf{.01}		&\textbf{.02}		&-	&-	&\textbf{.10}	&\textbf{.02}	&\textbf{.04}	&\textbf{.02}	&\textbf{.01}\\
uSRK P@RTP    	&.01		&-			&-		&.01		&-			&-		&.01	&-	&-	&.08	&.01	&.02	&-		&-\\
SRK R@RTP    	&\textbf{.03}	&\textbf{.03}	&\textbf{.02}	&\textbf{.05}	&-	&\textbf{.02}		&\textbf{.04}	&-	&-	&\textbf{.17}	&\textbf{.05}	&\textbf{.06}	&\textbf{.05}	&\textbf{.01}\\
uSRK R@RTP    	&(.01)		&.004		&-		&.01		&-			&-		&.01	&-	&-	&.14	&.03	&.03	&-		&-\\
SRK F1@RTP   	&\textbf{.01}	&\textbf{.01}	&-	&\textbf{.04}	&-	&\textbf{.01}		&\textbf{.03}	&-	&-	&\textbf{.13}	&\textbf{.05}	&\textbf{.05}	&\textbf{.03}	&\textbf{.01}\\
uSRK F1@RTP   	&(.01)		&-			&-		&.01		&-			&-		&.01	&-	&-	&.10	&.02	&.03	&-		&-\\
\hline
\end{tabular}
}
\end{table*}

\begin{table*}[t!]
\centering
\small
\caption{Comparing SemRe-Rank (SRK) and its unsupervised variant (uSRK, both with $z=200$) against each base ATE method. Only the changes over the base ATE methods are shown as points within a scale of [0, 1], and (brackets) indicate negative changes. \textbf{Bold} texts highlight the higher (if different) value between SRK and uSRK on each compared metric. }\label{tab_result_semrerank_full_z=200} 
\resizebox{150mm}{!}{
\begin{tabular}{|l |l |l |l|l |l|l|l|l|l|l|l|l|l|l|}
\hline
Metric     	&\rotatebox[origin=c]{270}{Basic}				&\rotatebox[origin=c]{270}{\shortstack{Combo\\Basic}}		&\rotatebox[origin=c]{270}{LP}				&\rotatebox[origin=c]{270}{NTM}	&\rotatebox[origin=c]{270}{PU}				&\rotatebox[origin=c]{270}{Vote\textsubscript{5}}	&\rotatebox[origin=c]{270}{CValue}	  &\rotatebox[origin=c]{270}{\shortstack{Gloss-\\Ex}} 	&\rotatebox[origin=c]{270}{RAKE}	&\rotatebox[origin=c]{270}{\shortstack{Rele-\\vance}}	&\rotatebox[origin=c]{270}{TFIDF}	&\rotatebox[origin=c]{270}{\shortstack{Weird-\\ness}}	&\rotatebox[origin=c]{270}{$\chi^{2}$}	&\rotatebox[origin=c]{270}{Vote\textsubscript{7}}\\
\hline
\multicolumn{15}{|c|}{ACLv2}\\
\hline
SRK P@K CNGs 	&\textbf{5}	&3		&2		&2				&\textbf{5}	&\textbf{4}		&\textbf{4}		&4				&5	&5				&\textbf{4}		&4			&\textbf{5}		&5\\
uSRK P@K CNGs 	&4			&3		&2		&\textbf{3}		&4			&3				&2				&\textbf{5}		&5	&5			&2				&4			&4				&5\\

SRK avg P@K	&\textbf{.024}		&\textbf{.014}	&\textbf{.01}			&\textbf{.016}	&\textbf{.03}		&\textbf{.01}	&\textbf{.024}	&\textbf{.012}	&.094			&\textbf{.118}	&\textbf{.068}	&\textbf{.03}	&\textbf{.042}	&.026\\
uSRK avg P@K	&.014	&.01	&.01	&.01	&.016	&-&.01	&.01	&.094	&.10	&.014	&.022	&.014	&.018\\
\hline
SRK P@RTP   &\textbf{.01}	&-		&-	&-		&\textbf{.01}	&-	&\textbf{.01}	&.01	&\textbf{.03}	&.02	&-	&-	&\textbf{.01}	&.01\\
uSRK P@RTP    	&-		&-	&-	&-		&-		&-	&-	&.01	&.02	&.02	&-		&-	&-		&\textbf{.02}\\

SRK R@RTP   &.003			&-		&-	&.04	&\textbf{.01}	&-	&\textbf{.002}	&.01	&\textbf{.04}	&.04	&-	&-	&\textbf{.01}	&.002\\
uSRK R@RTP    	&.003	&-	&-	&.04	&.005	&-	&-	&.01	&.02	&.04	&-		&-	&-		&\textbf{.01}\\

SRK F1@RTP  &\textbf{.01}	&-		&-	&.02	&\textbf{.01}	&-	&\textbf{.01}	&.01	&\textbf{.03}	&.03	&-	&-	&\textbf{.01}	&.01\\
uSRK F1@RTP   	&-		&-	&-	&.02	&.002	&-	&-	&.01	&.02	&.03	&-		&-	&-		&\textbf{.02}\\

\hline
\multicolumn{15}{|c|}{GENIA}\\
\hline

SRK P@K CNGs 	&4				&4		&4			&\textbf{5}		&2		&5		&5		&4		&5		&5		&\textbf{4}		&5		&5		&5\\
uSRK chng P@K  	&\textbf{5}		&4		&\textbf{5}	&4				&2		&5		&5		&4		&5		&5		&3				&5		&5		&5\\

SRK avg P@K		&\textbf{.012}	&.012	&.038			&\textbf{.01}	&\textbf{.032}	&\textbf{.01}	&\textbf{.02}	&.036	&\textbf{.062}	&\textbf{.12}	&\textbf{.026}			&.086	&\textbf{.026}	&\textbf{.01}\\
uSRK avg P@K	&.01			&.012	&\textbf{.044}	&.01			&.018			&.01			&.014			&.036			&.056			&.096			&.018					&.086	&.022			&.01\\
\hline
SRK P@RTP    	&\textbf{.01}	&-	&-	&-	&-	&-		&\textbf{.01}	&-		&.03	&\textbf{.01}	&.04	&.01	&.01	&.01\\
uSRK P@RTP    	&.004	&-	&-	&-	&-	&-		&-		&-		&.03	&-	&.04	&.01	&.01	&.01\\

SRK R@RTP    	&.01			&-	&-	&-	&-	&.004	&.01			&.007	&.04	&\textbf{.01}	&.04	&.01	&.01	&\textbf{.01}\\
uSRK R@RTP    	&.01	&-	&-	&-	&-	&.004	&.01	&.007	&.04	&-	&.04	&.01	&.01	&.004\\

SRK F1@RTP   	&.01	&-	&-	&-	&-	&-				&\textbf{.01}	&.003	&.03	&\textbf{.01}	&.04	&.01	&.01	&.01\\
uSRK F1@RTP   	&.01	&-	&-	&-	&-	&-		&.002	&.003	&.04	&-	&.04	&.01	&.01	&.01\\
\hline
\multicolumn{15}{|c|}{TTCm}\\
\hline
SRK P@K CNGs  	&2	&2	&\textbf{3}	&\textbf{4}			&\textbf{3}	&4	&3	&\textbf{3}	&\textbf{5}	&5	&\textbf{4}			&\textbf{5}	&\textbf{4}		&3\\
uSRK P@K CNGs  	&2	&2	&-	&1		&2		&4		&3		&2			&2		&5		&3		&4		&2		&\textbf{4}\\

SRK avg P@K	    &\textbf{.018}	&\textbf{.016}	&\textbf{.01}	&\textbf{.012}	&\textbf{.046}	&\textbf{.012}	&\textbf{.05}	&\textbf{.008}	&\textbf{.016}	&\textbf{.15}	&\textbf{.082}	&\textbf{.08}	&\textbf{.108}	&\textbf{.026}\\
uSRK avg P@K	&.012	&.01	&-	&.01	&-	&.01	&-	&-&.01	&.078	&.024	&.026	&0&.01\\

\hline
SRK P@RTP    	&-	&-	&\textbf{.01}	&\textbf{.05}	&-	&\textbf{.02}		&\textbf{.04}	&-	&\textbf{.01}	&\textbf{.24}	&\textbf{.05}	&\textbf{.11}	&\textbf{.02}	&\textbf{.04}\\
uSRK P@RTP    	&-		&-		&-		&.02	&-	&-		&-		&-		&-		&.08	&.03	&.06	&.01	&.01\\

SRK R@RTP    	&-				&-				&\textbf{.02}	&\textbf{.08}	&-	&\textbf{.03}		&\textbf{.08}	&\textbf{.004}	&\textbf{.02}	&\textbf{.35}	&\textbf{.09}	&\textbf{.14}	&\textbf{.03}	&\textbf{.08}\\
uSRK R@RTP    	&\textbf{.004}	&\textbf{.004}	&.01	&.02	&-	&-		&.01	&-		&.01	&.16	&.05	&.07	&.02	&.02\\

SRK F1@RTP   	&-	&-	&\textbf{.01}	&\textbf{.06}	&-	&\textbf{.02}		&\textbf{.05}	&-	&\textbf{.01}	&\textbf{.28}	&\textbf{.07}	&\textbf{.12}	&\textbf{.03}	&\textbf{.05}\\
uSRK F1@RTP   	&-		&-		&-		&.02	&-	&-		&.002	&-		&-		&.12	&.04	&.07	&.01	&.01\\

\hline
\multicolumn{15}{|c|}{TTCw}\\
\hline
SRK P@K CNGs  	&2	&1	&\textbf{2}	&\textbf{4}	&\textbf{2}	&1	&\textbf{3}			&\textbf{2}	&\textbf{1}	&\textbf{5}	&\textbf{4}	&\textbf{5}	&\textbf{4}	&\textbf{3}\\
uSRK P@K CNGs  		&2			&1		&-&2		&1			&1		&1		&1	&-&5		&1		&2		&1		&1\\

SRK avg P@K	    &.01	&.004	&\textbf{.01}	&\textbf{.03}	&\textbf{.034}	&\textbf{.01}	&.014	&\textbf{.01}	&\textbf{.004}	&\textbf{.098}	&\textbf{.048}	&.032	&\textbf{.04}	&\textbf{.038}\\
uSRK avg P@K	    &.01		&.004	&-&.01	&.01		&.01	&.01	&.01&-&.04	&.004	&.012	&.01	&.004\\

\hline
SRK P@RTP    	&\textbf{.01}	&.004	&\textbf{.01}	&\textbf{.03}	&\textbf{.034}	&\textbf{.01}		&\textbf{.014}		&\textbf{.01}	&\textbf{.004}	&\textbf{.098}	&\textbf{.048}	&\textbf{.032}	&\textbf{.04}	&\textbf{.038}\\
uSRK P@RTP    	&.006			&.004	&-				&.006			&.004			&.004				&.004				&.004			&-				&.04			&.004			&.012			&.008	&.004\\

SRK R@RTP    	&\textbf{.03}	&\textbf{.02}	&\textbf{.02}	&\textbf{.05}	&-&\textbf{.03}		&\textbf{.09}	&-&\textbf{.02}	&\textbf{.27}	&\textbf{.06}	&\textbf{.08}	&\textbf{.03}	&\textbf{.02}\\
uSRK R@RTP    	&.01			&.01		&-	&.01		&-		&.01	&(.01)&-&-&.09	&.02	&.02	&.01	&-\\

SRK F1@RTP   	&\textbf{.01}	&\textbf{.01}	&-&\textbf{.04}	&-&\textbf{.02}		&\textbf{.06}	&-&\textbf{.01}	&\textbf{.19}	&\textbf{.03}	&\textbf{.06}	&\textbf{.02}	&\textbf{.01}\\
uSRK F1@RTP   	&.003			&.003		&-	&.01		&-		&.01	&(.01)&-&-&.06	&.01	&.02	&.01	&-\\

\hline
\end{tabular}
}
\end{table*}

\section{Base ATE methods configurations} \label{appendix_config}
Both JATE 2.0 and ATR4S allow evaluating ATE methods in a uniform environment. This is achieved through using the same linguist processors to extract the same set of candidate terms for different ATE methods. While the two libraries do not support identical settings, we have ensured that they are as close as possible and that methods within each library use the same candidate term extraction process. 

Specifically, JATE 2.0 uses PoS sequence patterns to extract words and word sequences based on their PoS tags. The PoS patterns depend on different datasets. For GENIA and ACLv2, we use the same patterns as in \cite{Zhang2016}. For TTCw and TTCm, we use the patterns distributed with the datasets. We then process the candidates by removing leading and trailing stop words and non-alphanumeric characters, and only keep candidate terms that satisfy several conditions defined on: minimum character length (minc), maximum character length (maxc), minimum words (minw), and maximum words (maxw). 

ATR4S firstly extracts n-grams, then filters them by applying a generic PoS pattern and stop words removal. It also supports min/max char, and min/max word parameters. Table \ref{tab_ate_config} shows the details of the candidate term extraction configuration on all datasets. The slightly stricter constraints applied to both TTCw and TTCm datasets are used as a means to reduce incorrect candidate terms due to very sparse real terms in the datasets. Table \ref{tab_rtp} shows the number of candidate terms extracted from each dataset by each ATE method. Note that we do not use minimum frequency to filter candidate terms. Frequency based filtering is a common practice in ATE to reduce the number of false positives \cite{Zhang2016}, however, at the cost of losing true positives. Overall, Table \ref{tab_rtp} shows that the generic PoS patterns used by ATR4S generate more candidate terms on all datasets, while the domain specific PoS patterns used by JATE 2.0 capture more correct candidate terms (RTP). 

\begin{table}[t!]
\centering
\small
\caption{Configuration used by base ATE methods implemented in the ATR4S and the JATE 2.0 libraries. `N/A' indicates that the configuration parameter is not available for the implementation of that method.}\label{tab_ate_config} 



\begin{tabular}{l |l |l |l|l}
\hline
					&minc    &maxc	    &minw	&maxw \\
\hline
\multicolumn{5}{l}{Basic, ComboBasic, LP, NTM, PU (from ATR4S)} \\
\hline
GENIA     &2 	&N/A	&1	&5\\
ACLv2	  &2 	&N/A	&1	&5\\
TTCw  	  &3 	&N/A	&1	&4\\
TTCm 	  &3 	&N/A	&1	&4\\
\hline
\multicolumn{5}{l}{TFIDF, CValue, RAKE, Weirdness, Relevance, GlossEx, $\chi^{2}$ (from JATE 2.0)} \\
\hline
GENIA     &2 	&40	&1	&5\\
ACLv2	  &2 	&40	&1	&5\\
TTCw  	  &3 	&40	&1	&4\\
TTCm 	  &3 	&40	&1	&4\\
\hline
\end{tabular}
\end{table}

\end{appendices}
\end{document}